\documentclass[12pt,a4paper]{article}
\makeindex

\newcommand{\D}[2]{\frac{\partial #2}{\partial #1}}
\newcommand{\DD}[2]{\frac{\partial^2 #2}{\partial #1^2}}
\newcommand{\Dn}[3]{\frac{\partial^{#2} #3}{\partial #1^{#2}}}
\renewcommand{\vec}[1]{\mbox{\boldmath$#1$}}
\newcommand{\Ord}[1]{{\mathcal O}\left(#1\right)}
\newcommand{\etal}{\emph{et al}}
\newcommand{\avu}{{\bar u}}
\newcommand{\toe}{\textsc{toe}}
\newcommand{\ode}{\textsc{ode}}
\newcommand{\pde}{\textsc{pde}}
\newcommand{\sde}{\textsc{sde}}
\newcommand{\cD}{{\mathcal D}}
\newcommand{\cE}{{\mathcal E}}
\newcommand{\cG}{{\mathcal G}}
\newcommand{\cJ}{{\mathcal J}}
\newcommand{\cL}{{\mathcal L}}
\newcommand{\cM}{{\mathcal M}}
\newcommand{\cV}{{\mathcal V}}
\newcommand{\R}{{I\!\!R}}
\newtheorem{theorem}{Theorem}
\newcommand{\res}{\mbox{\sf R}}
\newcommand{\emidx}[1]{\emph{#1}\index{#1}}
\newcommand{\idx}[1]{#1\index{#1}}

\renewcommand{\sectionmark}[1]%
{\markright{\textnormal{\S\sf \thesection: #1}}}
\makeatletter\def\@oddfoot{\hfill\footnotesize\sf AJ Roberts, \today}\makeatother

\usepackage{epsfig}
\epsfxsize=\textwidth 
\renewcommand{\includegraphics}[1]{\epsfbox{#1}}

\begin{document}

\title{\sf Low-dimensional modelling \\ of dynamical systems}
\author{AJ Roberts\thanks{Dept.~Mathematics \& Computing, University
of Southern Queensland, Toowoomba, Queensland 4350, \textsc{Australia}.
E-mail: \texttt{aroberts@usq.edu.au}}}
\date{21 February, 1997}
\maketitle

\begin{abstract}
Consider briefly the equations of fluid dynamics---they describe the 
enormous wealth of detail in all the interacting physical elements of 
a fluid flow---whereas in applications we want to deal with a 
description of just that which is interesting.  In a wide variety of 
situations, simple approximate models are needed to perform practical 
simulations and make forecasts.  I review the \emph{derivation}, from 
a mathematical description of the detailed dynamics, of accurate, 
complete and useful low-dimensional models of the interesting dynamics 
in a system.  The development of centre manifold theory and associated 
techniques puts this modelling process on a firm basis.  As in 
Guckenheinmer \& Holmes \cite[\S2.5]{Guckenheimer83}:
\begin{quote}
	``\ldots these new methods will really be conventional 
	perturbation style analyses interpreted geometrically\ldots''
\end{quote}
But the geometric viewpoint of dynamical systems theory greatly 
enriches our approach by providing a rationale for also deriving 
correct initial conditions, forcing and boundary conditions for the 
models---all essential elements of a model.
\end{abstract}

\tableofcontents

\section{Introduction}
\label{Sintro}

In secondary school we learn of the parabolic flight of a ball.  In 
university physics we learn how it spins.  In elasticity courses we 
learn that a ball deforms as it spins and bounces.  In each successive 
stage of the modelling of the ball we deal with more details of the 
dynamics: first the position and velocity of the centre of mass, then 
position, orientation and their velocities, and lastly the position 
and velocity of every part of the ball.  \emph{In this paper I review 
the reverse of this process}: namely the reduction from equations 
describing the very detailed dynamics of a system down to simpler 
``model'' equations dealing with the dynamics at a relatively coarse 
level of description.

A model is ``simple'' if it deals with just a few characteristics of 
the physical problem.  For example, in the flight of the ball we 
describe just the movement of the centre of mass rather than the 
movement of each part of the ball.  Describing just a few 
characteristics and so written in terms of just a few parameters, 
\emph{a simple model is of low-dimension} when compared with the 
actual physical system.  A model is useful if it describes the 
dynamics of interest with very little extraneous details: we are 
usually interested in how the ball as a whole moves without needing to 
know anything about its deformation as it spins and bounces.  It is 
the rationale, theory and practice of such reduction in dimensionality 
that I review.

The general challenge is to start with an accurate and reliable 
description of the dynamics of a system of interest, a ``Theory Of 
Everything'' ({\toe}) as I wryly call it in this article, then to 
analyse it systematically and extract routinely the simple, 
low-dimensional dynamical models which are relevant in given 
situations:
\begin{displaymath}
	\mbox{{\toe}}\longrightarrow\mbox{model}\,.
\end{displaymath}
To do this, we invoke centre manifold theory, introduced in 
\S\ref{Sdimmod}, and associated techniques based upon geometric 
pictures in the state space of the dynamical system.  The use of 
centre manifold theory for this modelling was initiated by Coullet \& 
Spiegel \cite{Coullet83} and Carr \& Muncaster \cite{Carr83a,Carr83b}.  
The rationale of exponential collapse that underpins centre manifold 
theory has been also invoked in other methods, but I argue that this 
approach has many advantages over the alternatives.

Applications of the theory are widespread, some are discussed in 
\S\ref{Sappl} and an historical perspective given by Coullet \& 
Spiegel \cite[\S1]{Coullet83}.  Specific examples of this process of 
modelling by dimensional reduction are:
\begin{itemize}
\item elasticity $\longrightarrow$ rigid body motion (mentioned above); 

\item heat or mass transfer $\longrightarrow$ dispersion in a pipe or 
channel (\S\S\ref{SSdisp});

\item Navier-Stokes equations $\longrightarrow$ thin films, sheets \& 
jets of fluids (\S\S\ref{SSfilm});

\item Navier-Stokes \& heat $\longrightarrow$ convection (\S\S\ref{SSband}); 

\item Markov chain $\longrightarrow$ quasi-stationary distribution 
\cite{Pollett90}; 

\item elasticity $\longrightarrow$ beam theory of bending and torsion 
(\S\S\ref{SSbeam});

\item atmospheric models $\longrightarrow$ quasi-geostrophic 
approximation (\S\S\ref{SSgeost}).
\end{itemize}
Although my interest, and thus this review, lies primarily in the 
field of fluid mechanics, the modelling issues addressed are relevant 
to any evolutionary system.  However, there is not yet enough theory 
to support all the interesting applications of the concepts and 
techniques that we encounter.  The main limitation on rigour is that 
in many of these applications the ``low-dimensional'' model is still 
actually of ``infinite dimension'', a case for which there is almost 
no theory that is both rigorous and useful---more specific comments 
are sprinkled throughout.

There are many advantages in the centre manifold approach when 
compared with other methods of analysing dynamical systems to develop 
low-dimensional dynamical models.  

Competing small effects need not appear at leading order in the 
analysis, \S\ref{Ssmall}; consequently you obtain the flexibility to 
justifiably adapt the model to different applications \emph{without} 
redoing the whole analysis.  Just one example of this flexibility is 
used to modify the governing equations of thin fluid films, 
\S\S\ref{SSfilm}, so that the model incorporates the extra dynamical 
degree of freedom to resolve wave dynamics on the film.  To recover a 
description of the physical problem, we sum to high-order in the 
modification while only considering adequate low-order physical terms.  
Such flexibility also justifies a local description of the interaction 
between counter-propagating waves.

The geometric picture of evolution near the centre manifold suggests 
analysis, \S\S\ref{SSprojic}, that provides correct initial conditions 
to use with the model to make correct long-term forecasts.  The 
algebra is based upon how neighbouring trajectories evolve, 
identifying which ones approach each other exponentially quickly and 
thus have the same long-term evolution.  For example, in the long-term 
dispersion down a channel (\S\S\ref{SSdisp}) we can discern the 
difference between dumping contaminant into the slow moving flow near 
the bank and into the fast core flow.  Normal form transformations, 
\S\S\ref{SSnorm}, also illustrate the projection of initial conditions 
onto the centre manifold.  In addition, normal form transformations 
show limitations in the so-called slow manifold models, \S\ref{Sslow}, 
that are formed by removing the dynamics of fast oscillations (as in 
beam theory or rigid-body dynamics).  Through nonlinear interaction, 
neglected fast waves may resonate and cause inevitable errors to 
accumulate over time.  In contrast, centre manifold models guarantee 
the existence of forecasts that are accurate exponentially quick.

One attribute of basic centre manifold theory is that it deals with 
autonomous dynamical systems.  However, in the presence of a 
time-dependent forcing, \S\S\ref{SSdforc}, the system is pushed away 
from the centre manifold and so we use the geometric projection of 
initial conditions to determine what forcing is appropriate in the 
model.  I show, for example, that a model may be very sensitive to a 
forcing which more primitive approaches would neglect.  There is also 
many interesting issues in the modelling of noisy dynamical systems, 
as expressed by stochastic differential equations.  I argue, 
\S\S\ref{SSnoise}, that centre manifold concepts provide relatively 
straightforward tools to approach this problem rationally.  These 
methods should apply to interesting questions such as: what influence 
may turbulence have on dispersion?  and how does substrate roughness 
affect the flow of thin films?

Many useful models are expressed in terms of partial differential 
equations in space and time.  Such partial differential equations must 
have boundary conditions.  For example, models for dispersion in a 
channel or for beam theory require boundary conditions at the inlet 
and outlet of the channel or the ends of the beam respectively.  
Arguments based upon the \emph{spatial evolution} away from the 
boundary, applied to both the full system and the model, give a 
rationale which provides correct boundary conditions, 
\S\ref{Sbound}.  To leading order these boundary conditions are 
typically those obtained from physical heuristics.  However, there 
are corrections accounting for more subtle features of the dynamics.  

Computing details of the centre manifold and its dynamic model often 
involves considerable algebra.  This is especially true for centre 
manifolds of more than just a few dimensions as the algebraic 
complexity of the model may increase combinatorially.  Computer 
algebra can be used to minimise the human labour involved.  After all:
\begin{quote}
	``It is unworthy of excellent persons to lose hours 
	like slaves in the labour of calculation''\ldots Gottfried Wilhelm von 
	Leibniz.
\end{quote}
The challenge addressed in \S\ref{Scompalg} is to develop algorithms 
for computer based algebra packages which are simple and reliable to 
implement.

All the above aspects are features of a complete approach to modelling 
dynamical systems based upon centre manifold theory, techniques and 
concepts.  

Throughout this review we focus on continuous time dynamics, 
flows, as expressed through differential equations.  Similar concepts 
and analysis can be also developed for discrete maps but I will not 
elaborate on these.

\section{Rational modelling theory}
\label{Sdimmod}

Simple models must have low dimension corresponding to the few 
variables in the model.  Since a model must have something to say 
about the \toe{}---states of the model correspond to states of the 
\toe---the state space of the model must be able to be embedded in the 
high-dimensional state space of the \toe{}.  Typically we imagine that 
the state space of a model forms a low-dimensional differentiable 
manifold within the state space of the \toe.\index{dimensionality}

If the model is to accurately capture the dynamics of the {\toe}, then 
the manifold must be made of some of the trajectories of the {\toe}.  
The detail lost in apparently ignoring all the dynamics outside the 
states described by the model is an inevitable consequence of forming 
a simple, low-dimensional model.  The process of analysing the \toe{} 
and creating a low-dimensional model is sometimes termed \emidx{coarse 
graining} \cite[e.g.]{Muncaster83b} because of the loss of fine detail 
in forming a model.  In this section I introduce centre manifold 
theory as a basis for the \idx{rational modelling} of dynamical 
systems, as first recognised by Coullet \& Spiegel \cite{Coullet83} 
and Carr \& Muncaster \cite{Carr83a,Carr83b}.

\subsection{Exponential collapse gives a rationale}
\label{SSrat}

\index{exponential collapse}
If many modes of the {\toe} decay exponentially, then all that is left 
after the transients decay are the \emph{relatively} slowly evolving 
modes of long-term importance.  The evolution of these few significant 
modes effectively forms a low-dimensional dynamical system on a 
low-dimensional set of states in state space.  Through the rapid 
exponential decay all neighbouring trajectories are quickly attracted 
to these low-dimensional dynamics and so they form an accurate 
low-dimensional model of the \toe.

This idea also lies behind the construction of so-called 
\emidx{inertial manifolds} by Temam \cite{Temam90} and others.  
Functional analysis is used to construct inertial manifolds and 
estimate some properties such as their dimension (see Foias \etal{} 
\cite{Foias85a,Foias88a} for example).  But what if we are not just 
interested in the ultimate attractor, but are curious about 
long-lasting transients?  After all, the eventual fate of the universe 
is either the big crunch or a featureless death by high entropy---but 
before these become an overriding issue to humans there are many 
problems of interest to study, albeit transient.  A more prosaic 
example is the dispersion of material in an infinitely long channel or 
pipe: the ultimate attractor is a completely dispersed contaminant of 
effectively zero concentration everywhere; however, as seen in 
\S\S\ref{SSdisp}, we are very interested in modelling the long 
transient of how a contaminant spreads as it is carried downstream.

One case where the distinction between exponential decay and 
long-lasting importance can be made with absolutely clarity is in the 
neighbourhood of an equilibrium or fixed point.  Without loss of 
generality we take the reference equilibrium at the origin.  Let the 
\emph{linearised} dynamics be $\dot{\vec u}=\cL\vec u$, then the 
eigenvalues of the linear operator $\cL$ determine the dynamics in the 
neighbourhood:
\begin{itemize}
\item  modes with $\Re(\lambda)<0$ decay exponentially;  
	
\item modes with $\Re(\lambda)=0$ do not decay, are long-lasting and 
form the basis of a low-dimensional model.
\end{itemize}
An elementary example from \cite{Roberts85b} is
\begin{equation} 
\dot x=-xy \,, 
\quad\mbox{and}\quad
\dot y=-y+x^2-2y^2 \,. \label{Eegcm}
\end{equation}
\begin{figure}[tbp]
	\centerline{\includegraphics{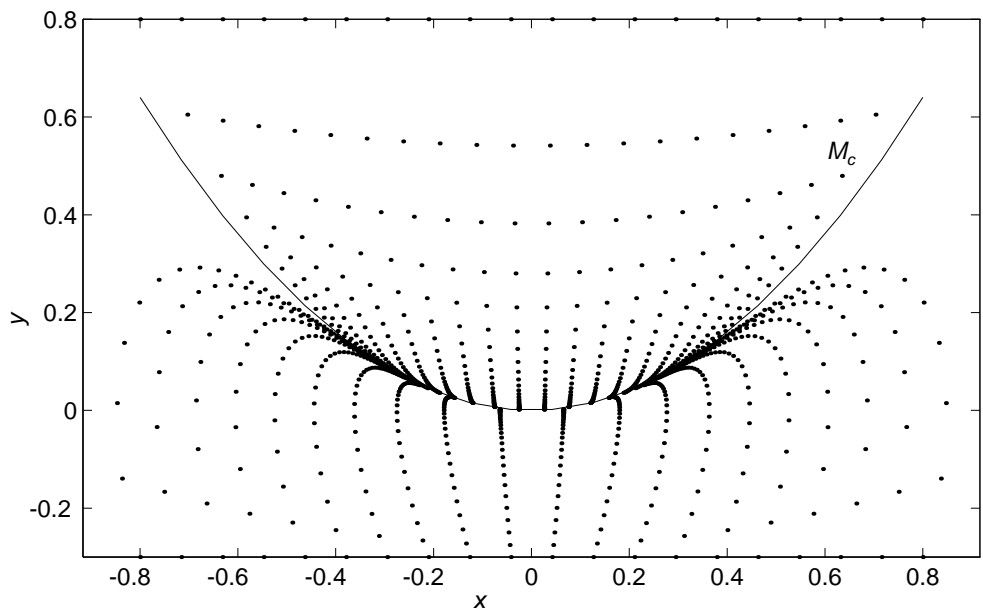}}
	\caption{trajectories of the dynamical system 
	(\protect\ref{Eegcm}) plotted every $\Delta t=0.2$ to show the 
	rapid approach, roughly in 2 units of time, to the centre 
	manifold, $\cM_c$.  Observe how all these initial conditions 
	collapse onto a one-dimensional set of states in which \emph{all} the 
	long-term dynamics take place.}
	\protect\label{Fegcm}
\end{figure}%
Very quickly all trajectories approach a curved ``subspace'' in state 
space, called the centre manifold, $y=x^2$ in this example as shown in 
Figure~\ref{Fegcm}.  Thus no matter what the initial condition, the 
only states of long-term interest are those of the centre manifold.  
The evolution on the centre manifold, here $\dot x=-x^3$, then forms a 
low-dimensional model of \emph{all} the dynamics in the system.

In the above, I introduced the criterion that $\Re(\lambda)=0$ for 
determining the modes that form the low-dimensional model.  Arguments 
have been devised to relax this restrictive equality.  An argument 
could be made that modes with $\Re(\lambda)<-\gamma<0$ decay 
exponentially, whereas modes with $\Re(\lambda)\geq-\gamma$ are at 
least \emph{longer}-lasting and form the basis of a low-dimensional 
model---here it is the modes with $\Re(\lambda)<-\gamma$ that 
``collapse'' the state space.  For an elementary example, also from 
\cite{Roberts85b}, consider
\begin{equation} 
\dot x=\epsilon x -xy \,, 
\quad\mbox{and}\quad
\dot y=-y+x^2-2y^2 \,, \label{Eegcum}
\end{equation}
\begin{figure}[tbp]
	\centerline{\includegraphics{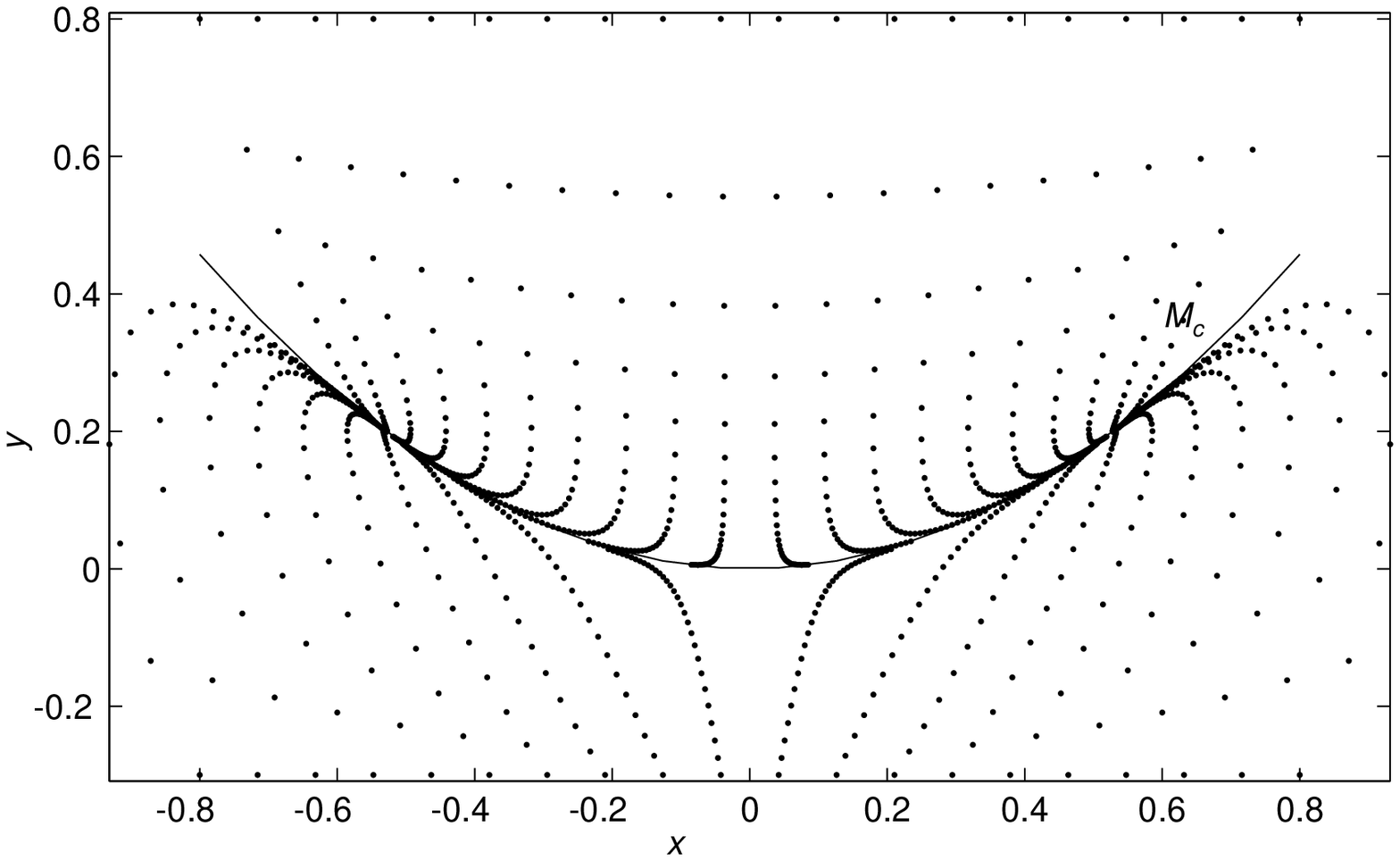}}
	\caption{trajectories of the dynamical system 
	(\protect\ref{Eegcum}) with parameter 
	$\epsilon=0.2$ plotted every $\Delta t=0.2$ to show the rapid 
	approach, roughly in 2 units of time, to the centre-unstable 
	manifold, $\cM_{cu}$.  Again observe the 
	collapse onto a one-dimensional set of states.}
	\protect\label{Fegcum}
\end{figure}%
which has the exponentially attractive manifold 
\begin{equation}
	y=\frac{x^2}{1+2\epsilon}\,,
	\label{Eegcu}
\end{equation}
as seen in Figure~\ref{Fegcum}, on which the model evolution is $\dot 
x=\epsilon x-x^3/(1+2\epsilon)$.  For $\epsilon\geq0$ this is termed a 
\emidx{centre-unstable manifold}, a concept used by Armbruster \etal{} 
\cite{Armbruster89} to investigate the Kuramoto-Sivashinksy dynamics, 
by Cheng \& Chang \cite{Cheng92} for subharmonic instabilities of 
waves, and by Chow \& Lu \cite{Chow88} to compare with the method of 
averaging.  For general $\epsilon$, not too large in magnitude, 
(\ref{Eegcu}) describes an \emidx{invariant manifold} 
\cite[\S1.1C]{Wiggins90}, which when exponentially attractive may be 
used to create accurate models \cite{Roberts89,Roberts90}, such as 
that of dispersion in simple channels \cite{Watt94b}.  Invariant 
manifolds, based upon modes with $\Re(\lambda)>-\gamma$, may improve 
the numerical solution of spatio-temporal \pde{}s through what others 
have called the \emidx{nonlinear Galerkin method} 
\cite{Marion89,Temam89,Foias88b,Luskin89,Foias94}.  The disadvantage 
of these more general invariant manifolds for modelling is that the 
consequent algebraic analysis in constructing the manifold is 
significantly more difficult.  In \S\S\ref{SSextend} I discuss a 
method to approximate these more general invariant manifold models 
while maintaining the relatively simple algebra associated with centre 
manifolds.

A quite different concept in modelling dynamics is what van Kampen 
\cite{vanKampen85} calls that of the \emidx{guiding centre}.  In the 
presence of fast oscillations, such as short-period waves, we may be 
only interested in the long-term evolution of the dynamics of the 
mean.  For example, incompressible fluid dynamics ignores fast sound 
waves (for example, see the quasi-incompressible approximation of 
Durran \cite{Durran89}).  In such a case the particular dynamical flow 
without oscillations acts as a centre for flow with rapid oscillations 
and so is considered to form a low-dimensional model.  In many 
applications, such as beam theory \cite{Roberts93} or atmospheric 
flows \cite[e.g.]{Lorenz86}, the model dynamics are known as the slow 
dynamics on the \emidx{slow manifold}.  However, the modelling issues 
associated with this concept are much more delicate and I discuss them 
in more detail in \S\ref{Sslow}.  For the moment let us concentrate upon 
exponential collapse to the centre manifold as a rationale for forming 
low-dimensional dynamical models.

A completely different approach, albeit justified by the same 
exponential collapse, is to use \idx{symmetry} considerations to 
sketch out possible structurally stable, low-dimensional models.  See 
\cite{Edwards94} for example, or the review by Crawford \& Knobloch 
\cite{Crawford91} on symmetry in fluid dynamics.  The limitation of 
this approach is that one can practically never determine quantitative 
coefficients in the model.  Hence such an approach may be satisfactory 
for predicting the broad range of phenomena possible, but it cannot 
produce a model able to make detailed forecasts about a specific 
physical situation.

\subsection{Centre manifolds}

The centre manifold, $\cM_c$, is the curved ``subspace'' to which all 
trajectories in the neighbourhood of a fixed point are attracted 
exponentially quickly.  Being composed of trajectories of the {\toe} 
and being low-dimensional, the evolution on $\cM_c$ qualifies as 
forming a model of the {\toe}.

Three theorems claim:
\begin{enumerate}
\item $\cM_c$ \emph{exists} for the previously mentioned structure of 
eigenvalues, provided the nonlinear terms are not ``badly'' behaved;
 
\item $\cM_c$ is \emph{relevant} as all solutions in the neighbourhood 
are attracted exponentially to a solution on $\cM_c$; 

\item $\cM_c$ may be \emph{constructed} to any desired degree of 
accuracy (asymptotically speaking).
\end{enumerate}
Detailed proofs of these theorems have, for example, been given in the 
excellent little book by Carr \cite{Carr81}, and more recently by 
Vanderbauwhede and Iooss \cite{Vanderbauwhede88,Vanderbauwhede89}.  
In this subsection I address their role in low-dimensional modelling.

\subsubsection{Existence}
\label{SSSexist}

Most statements of theory address dynamical systems in the 
\emidx{separated form}:
\begin{eqnarray}
	\dot{\vec x} & = & A\vec x+\vec f(\vec x,\vec y)\,,
	\label{Estdfmx}  \\
	\dot{\vec y} & = & B\vec y+\vec g(\vec x,\vec y)\,,
	\label{Estdfmy}
\end{eqnarray}
where the eigenvalues of the matrix $A$ have real-part zero, the 
eigenvalues of the matrix $B$ have strictly negative real-part, and 
$\vec f$ and $\vec g$ are quadratically nonlinear functions at the 
origin, of $\Ord{x^2+y^2}$ where $x=|\vec x|$ and $y=|\vec y|$.  
However, we address dynamical systems, the \toe, in the general form
\begin{equation}
	\dot{\vec u}=\cL\vec u+\vec f(\vec u)\,,
	\label{Egenfm}
\end{equation}
where $\vec u(t)\in\R^n$, the linear operator $\cL$ has $m$ 
eigenvalues with zero real-part (counted according to their 
multiplicity), the remaining eigenvalues are strictly negative and 
bounded above by $-\gamma<0$, and $\vec f$ is a quadratically 
nonlinear at the origin, of $\Ord{u^2}$.  A linear transformation of 
variables will in theory transform the dynamical system from one form 
to the other.  However, in practise we prefer to perform analysis with 
meaningful physical variables and so the general form~(\ref{Egenfm}) 
is more appropriate in applications.

\begin{figure}[tbp]
	\centerline{\includegraphics{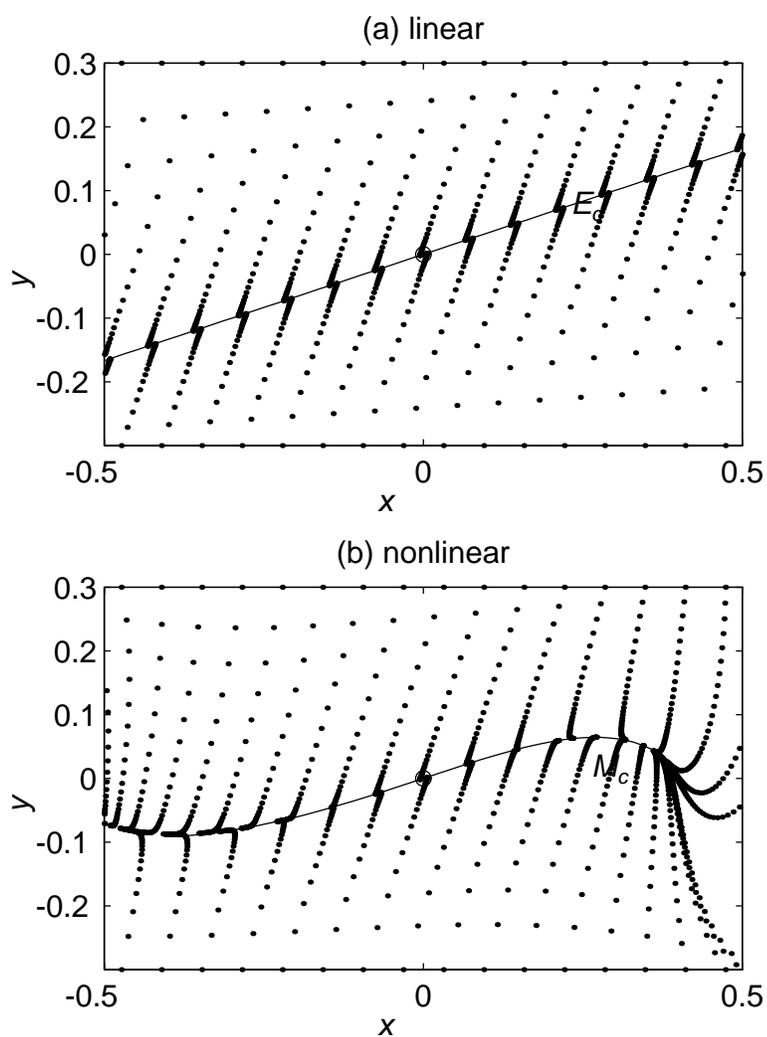}}
	\caption{shows an example of the exponential collapse of (a) 
	linear dynamics onto the centre subspace $\cE_c$, each dot on a 
	trajectory a fixed time $\Delta t$ apart, to compare with (b) 
	showing that the nonlinear dynamics ``bend'' the subspace to the 
	centre manifold $\cM_c$ and produce a slow evolution thereon.}
	\protect\label{Fnonlin}
\end{figure}

Linearly, according to $\dot{\vec u}=\cL\vec u$, all the dynamics will 
collapse exponentially\index{exponential collapse} quickly onto the 
\emidx{centre subspace},
\begin{displaymath}
	\cE_c=\mbox{span}\left\{\vec e_1,\ldots,\vec e_m\right\}\,,
\end{displaymath}
where $\vec e_j$ are the $m$ eigenvectors (or generalised 
eigenvectors) of the $m$ eigenvalues with zero real-part (counted 
according to multiplicity).  Theory asserts that at least near the 
origin the nonlinear terms just ``bend'' this invariant subspace and 
modify the dynamics on the subspace, as shown in Figure~\ref{Fnonlin}.

\begin{theorem}[existence]
\label{Tcmexists}
sufficiently near the origin, in some neighbourhood $U$, there exists 
an $m$-di\-men\-sion\-al invariant manifold for~(\ref{Egenfm}), 
$\cM_c$, with tangent space $\cE_c$ at the origin, in the form $\vec 
u=\vec v(\vec s)$ (that is, locations on $\cM_c$ are parameterised by 
a set of variables $\vec s$, often called the \emidx{order 
parameters}).  The flow on $\cM_c$ is governed by the $m$-dimensional 
dynamical system
\begin{equation} 
\dot{\vec s}=\cG\vec s+\vec g\left(\vec s\right) \,, \label{Egenm} 
\end{equation} 
where $\cG$ is the restriction of $\cL$ to $\cE_c$, and the nonlinear 
function $\vec g$ is determined from $\cM_c$ and~(\ref{Egenfm}).  
Because of the nature of the eigenvalues of $\cL$ this invariant 
manifold is called a \emidx{centre manifold} of the system.
\end{theorem}

A centre manifold is, in some neighbourhood of the origin, at least 
as differentiable as the nonlinear terms $\vec f$.  However, it may 
not be analytic even though $\vec f$ is analytic.  Also, a centre 
manifold need not be unique, but the differences between the possible 
centre manifolds are of the same order as the differences we set out 
to ignore in establishing the low-dimensional model 
\cite{Roberts85b,Roberts89}.  Non-uniqueness, when it arises, is 
irrelevant for modelling.\index{non-uniqueness}

In application, such as the fluid dynamics problems discussed in 
\S\ref{Sappl}, we need theory dealing with not only infinite 
dimensional \toe's, but also infinite dimensional centre manifolds.  
Extensions of the above theorem to infinite dimensional dynamics are 
extant but limited.  The most general theory I am aware of is 
currently due to Gallay \cite{Gallay93}, but it suffers from 
restrictions upon the nonlinear terms, they have to be bounded, which 
limits its rigorous application.  Scarpellini \cite{92i:58170} 
apparently places significantly less restrictions upon the 
nonlinearities in the dynamical equations, but while he addresses 
infinite dimensional centre manifolds, the results are severely 
constrained by requiring finite dimensional stable dynamics.  
H\u{a}r\u{a}gus \cite{Haragus95,Haragus95b} has developed theory 
supporting \idx{infinite dimensional models}, such as the 
\idx{Korteweg-de Vries equation}, but only by placing extremely 
limiting restrictions upon the time derivatives in the \toe.

\subsubsection{Relevance}

\index{relevance}
Based on the rationale of neglecting rapidly decaying transients, our 
intention is to consider~(\ref{Egenm}) as a simple model system for 
the \toe~(\ref{Egenfm}); simpler in the sense that it has lower 
dimensionality, $m$ instead of $n$.  However, there is a subtle point 
which is often overlooked.  We must be assured that the actual 
solutions of the model~(\ref{Egenm}) do indeed correspond to solutions 
of the full system~(\ref{Egenfm}).  Exponential attraction to an 
invariant manifold is \emph{not} enough to assure us that the 
solutions on the manifold actually model accurately all the nearby 
dynamics.\footnote{For example, consider the system $\dot x=x+xy$ and 
$\dot y=-y$ which has the exponentially attractive unstable manifold 
$y=0$.  There is generally an unavoidable $\Ord{1}$ discrepancy 
between the full system, with solutions $x(t)\sim Ae^t+B$, and the 
low-dimensional model, $\dot x=x$ with $x(t)\sim Ae^t$, despite the 
exponential collapse onto the unstable manifold.}

\begin{theorem}[relevance]
\label{Tcmrel}
The neighbourhood $U$ may be chosen so that all solutions 
of~(\ref{Egenfm}) staying in $U$ tend exponentially to some solution 
of~(\ref{Egenm}).  That is, for all solutions $\vec u(t)\in U$ for all 
$t\geq 0$ there exists a solution $\vec s(t)$ of the 
model~(\ref{Egenm}) such that
		\begin{displaymath}
			\vec u(t)=\vec v\left(\vec 
			s(t)\right)+\Ord{e^{-\gamma' t}} \quad\mbox{as}\quad 
			t\to\infty\,,
		\end{displaymath}
		where $-\gamma'$ may be \emph{estimated} as 
		$-\gamma$, the upper bound on the negative eigenvalues of the 
		linear operator $\cL$.
\end{theorem}

This theorem is crucial to modelling; it asserts that for a wide 
variety of initial conditions the dynamics of the \toe{} decays 
exponentially quickly to a solution which can be predicted by the 
low-dimensional model.  Somewhat pessimistically, the theorem requires 
initial conditions to be sufficiently small, namely in $U$.  But 
consider the system shown in Figure~\ref{Fegcm}.  Observe that all 
trajectories within the picture asymptote exponentially to the centre 
manifold.  Thus the conclusions of this theorem are correct for at 
least \emph{all} initial conditions within the figure.  In practice 
the neighbourhood $U$ may be quite large.

\subsubsection{Approximation}
\label{SSSapprox}

We need to find an equation to solve which gives the centre manifold 
$\cM_c$.  This is obtained straightforwardly by substituting the assumed 
functional relations, that $\vec u(t)=\vec v(\vec s(t))$ where 
$\dot{\vec s}=\cG\vec s+\vec g(\vec s)$, into the \toe~(\ref{Egenfm}) 
and using the chain rule for time derivatives to obtain
\begin{equation}
		\cL\vec v(\vec s)+\vec f\left(\vec v(\vec s)\right)
    =\D{\vec v}{\vec s}\left[\cG\vec s+\vec g\left(\vec s\right)\right]\,.
\label{Ecmei}
\end{equation}
This is the equation to be solved for the centre manifold $\cM_c$. 

An extra condition is that $\cE_c$ is the tangent space of $\cM_c$ at 
the origin.  Put more crudely, this requires that $\vec v$ is 
quadratically close to $\cE_c$.  This condition ensures that the 
constructed manifold truly contains the whole of the critical centre 
modes, and nothing but the centre modes.  Without it, the solution 
of~(\ref{Ecmei}) could be based on an almost arbitrary mixture of 
linear modes.  Indeed, other invariant manifolds of note 
satisfy~(\ref{Ecmei}) but are tangent to different subspaces as $\vec 
s\to 0$.

It is typically impossible to find exact solutions to~(\ref{Ecmei}). However, 
in applications we may approximate $\cM_c$ to any desired accuracy. 
For functions $\vec\phi:\R^m\to\R^n$ (imagine that $\vec\phi$ approximates 
the shape $\vec v$) and $\vec\psi:\R^m\to\R^m$ (imagine that $\vec\psi$ 
approximates the evolution $\dot{\vec s}=\cG\vec s+\vec g(\vec s)$) define the 
\emidx{residual} of~(\ref{Ecmei})
\begin{equation}
	        \res(\vec\phi,\vec\psi)=\D{\vec s}{\vec\phi}\vec\psi(\vec s)
	        -\cL\vec \phi(\vec s)
	        -\vec f\left(\vec \phi(\vec s)\right) \,,
	        \label{Erestoe}
\end{equation}
and observe that $\cM_c$ satisfies $\res(\vec v(\vec s),\cG\vec s+\vec g(\vec 
s))=\vec 0$.
\begin{theorem}[approximation]
\label{Tcmapprox}
    If the tangent space of $\vec \phi(\vec s)$ at the origin is $\cE_c$,
    and the residual $\res(\vec\phi,\vec\psi)=\Ord{s^p}$ as $ s\to 0$
    for some $p>1$ (where $s=|\vec s|$) then $\vec v(\vec s)=\vec\phi(\vec s)
    +\Ord{s^p}$ and $\dot{\vec s}=\cG\vec s+\vec g(\vec s)=\vec\psi(\vec s) 
    +\Ord{s^p}$ as $s\to 0$.
\end{theorem}
That is, if we can satisfy~(\ref{Ecmei}) to some \idx{order of accuracy} 
then the centre manifold is given to the same order of accuracy.

In problems specified in the separated 
form~(\ref{Estdfmx}--\ref{Estdfmy}) the centre manifold may be 
determined simply by iteration \cite[e.g.]{Carr81}.  However, 
typically a solution is sought in the form of an asymptotic power 
series in $\vec s$ as developed by Coullet \& Spiegel \cite{Coullet83} 
for the general form~(\ref{Egenfm}) (and reinvented by Leen 
\cite{94a:58169}).  Such a power series solution looks very like the 
\idx{Lyapunov-Schmidt method} \cite[e.g.]{93e:58047} for determining the 
nontrivial fixed points near a simple bifurcation; in contrast though, 
centre manifold theory also determines the dynamics, see 
\cite{95k:76046,93e:35105,92d:58035,88m:58134} for comparisons.  
Recently, I developed \cite{Roberts96a} an iterative algorithm, for 
the general form, with the virtue that it is readily implemented in 
computer algebra, see \S\ref{Scompalg}.

Centre manifold models need not be anchored to just the one 
equilibrium.  Some very interesting models are constructed for the 
dynamics in the neighbourhood of a \idx{manifold of equilibria}.  Such 
models may be uniformly valid across the whole set of equilibria.  One 
example is given in \S\S\ref{SSfilm} where a fluid at rest of constant 
but \emph{arbitrary} thickness provides the reference equilibria for a 
model of the evolution of the fluid film's thickness.

\subsection{A vital extension}
\label{SSextend}

Requiring that $\Re(\lambda)=0$ precisely as the criterion for the 
modes of the dynamical model is far too restrictive in practice.  A 
trick rescues the theory from obscurity.

It is easiest to see the trick in an example: to~(\ref{Eegcum}) adjoin 
the trivial dynamical equation $\dot \epsilon=0$.  Based on the 
linearisation about the equilibrium at the origin in the $\epsilon 
xy$-space (note that in this space $\epsilon x$ is a nonlinear term), 
there then exists a two-dimensional centre manifold, parameterised by 
$\epsilon$ and $x$.  The relevance and approximation theorems may 
similarly be applied to construct the centre manifold for small $x$ 
and $\epsilon$, and to validate the model, that
\begin{displaymath}
	\dot\epsilon=0\,,
	\quad\mbox{and}\quad
	\dot x=\epsilon x-x^3+\Ord{s^4}\,,
\end{displaymath}
where $s=|(\epsilon,x)|$.  Fixing upon any small value for $\epsilon$ 
then gives a model involving only the evolution of $x$, from which 
one may predict, for example, the presence of a \idx{pitchfork 
bifurcation}.

This example shows that we may apply centre manifold theory to 
problems which do not precisely fit the linear structure required by 
the theory.  By adjoining trivial dynamical equations we may, in 
essence, perturb eigenvalues with small real-part, either negative or 
positive, so that the ``interesting'' modes then have eigenvalues that 
precisely satisfy $\Re(\lambda)=0$ and hence are included in the 
centre manifold.  This trick can be not only used to unfold 
bifurcations, as above, it may also be used to partially justify the 
long-wave approximation (\S\S\ref{SSdisp}), and to approximate 
``hard'' problems by writing them as perturbations of easier problems 
(\S\S\ref{SSfilm}).

Because \idx{parameters} are so important in application, it is useful to 
quote a more powerful and flexible approximation theorem.  Consider 
the extended dynamical system with $\ell$ parameters $\vec\epsilon$:
\begin{equation}
    \dot{\vec\epsilon}=\vec 0\,,
    \quad\mbox{and}\quad
	\dot{\vec u}=\cL\vec u+\vec f(\vec\epsilon,\vec u)\,,
	\label{Egenp}
\end{equation}
with $\cL$ as before.  For functions 
$\vec\phi:\R^\ell\times\R^m\to\R^n$ approximating the centre manifold, 
and $\vec\psi:\R^\ell\times\R^m\to\R^m$ approximating the evolution 
thereon, one may argue that
\begin{theorem}[parameter approximation]
\label{Tcmparam}
	If the tangent space of $\vec \phi(\vec\epsilon,\vec s)$ at the 
	origin is $\R^\ell\times\cE_c$, and the residual of~(\ref{Egenp}) 
	is $\res(\vec\phi,\vec\psi)=\Ord{s^p+\epsilon^q}$ as 
	$(\epsilon,s)\to \vec 0$ for some $p>1$ and $q\geq1$, then $\vec 
	v(\vec\epsilon,\vec s)=\vec\phi(\vec\epsilon,\vec s) 
	+\Ord{s^p+\epsilon^q}$ and $\dot{\vec s}=\cG\vec s+\vec 
	g(\vec\epsilon,\vec s)=\vec\psi(\vec\epsilon,\vec s) 
	+\Ord{s^p+\epsilon^q}$ as $(\epsilon,s)\to \vec 0$.  Arguments also 
	justify the same statement but with errors $\Ord{s^p,\epsilon^q}$.
\end{theorem}
Consequences of this property are followed up in \S\ref{Ssmall} where I 
discuss the flexibility of centre manifold models.

\section{Physical applications show the potential}
\label{Sappl}

Centre manifold analysis has often been applied to reduce dynamics 
down to a ``handful'' of \ode{}s, typically the centre manifold is of 
dimension 1--3.  Also typically, a normal form transformation is then 
used to classify the dynamics obtained on the centre manifold.  This 
established approach is described in various books dealing with 
dynamical systems, see Carr \cite{Carr81}, Guckenheimer \& Holmes 
\cite[Ch.3]{Guckenheimer83}, Wiggins \cite[Ch.2]{Wiggins90}, Iooss \& 
Adelmeyer \cite[Ch.1]{Iooss92}, or Kuznetsov 
\cite[Ch.3--5]{Kuznetsov95}.  However, there is much greater scope for 
the use of centre manifold analysis in low-dimensional modelling.  
After a brief dash through some standard applications of the theory, 
in the rest of this review I outline some of the interesting 
applications that extend beyond current rigorous theory.

\subsection{A dash of straightforward applications}
\index{standard applications}

\begin{itemize}
\item Fluid mechanics\index{fluid mechanics} is a haven of nonlinear 
dynamics and the application of centre manifold theory.  The theory 
has been used in analysing the dynamics of Taylor vortices in 
Taylor-Couette flow \cite{89i:76002,Iooss92,Iooss92}, and the 
non-axisymmetric dynamics \cite{89m:35176,93d:58111} involving mode 
competition.  Mode interactions in rotating convection 
\cite{91g:76038} are analysed with centre manifolds, as are the 
dynamics of convection in porous media by Neel 
\cite{92b:76088,93h:76073,95h:76110} and others \cite{92k:76032}.  
Arneodo \etal{} \cite{Arneodo85a,Arneodo85b,Arneodo85c,95k:76046} 
reduced the dynamics of triple convection down to a set of three 
coupled \ode{}s, numerically verified the modelling and then proved 
the existence of chaos.

\item A number of studies 
\cite[e.g.]{89i:93086,Chen92,92b:93060,94h:34082b} have used centre 
manifold theory to help analyse the effectiveness of \idx{feedback control} 
in nonlinear dynamics \cite{90m:93068}, sometimes the centre manifold 
is called the slow manifold \cite{91i:93070,91d:73049,89j:93047}, see 
\S\ref{Sslow}.  These control problems are often concerned with 
systems where the control is delayed, leading to formally 
infinite-dimensional dynamics, and they investigate the effect of the 
delay on the dynamics through a low-dimensional centre manifold model 
\cite[e.g.]{Boe89,92d:58129,95m:34124}.  Some studies examine the 
centre manifolds of delay differential equations in their own right, 
for example: existence in \pde{}s \cite{93j:34116}, and the effects of 
multiple delays \cite{92m:34156,95j:34100}.  Usually, in these 
applications any finite time delay is washed out by the slow-evolution 
paradigm on the centre manifold.  However, in some other applications 
(see \S\S\ref{SSdisp}) it is attractive to transform a centre manifold 
model into an equivalent model which possesses delays or ``memory.''

\item Other mechanical systems of interest involve nonlinear elastic 
\index{nonlinear elasticity} \cite{95k:73075,94h:34082b} and 
viscoelastic \cite{89d:70022,95f:73059} springs.  The usual aim is to 
determine the onset and form of nonlinear oscillations, such as 
airfoil flutter \cite{95k:73075} or chaos in pipes conveying fluid 
\cite{94j:73034}, near some reference equilibrium.
	
\item The field of \idx{economics} is increasingly using the concepts and 
techniques of nonlinear dynamics, as described by Chiarella 
\cite{92i:90026}, and centre manifold theory has a role to play 
\cite[e.g.]{95j:90017}.

\item A remarkable application occurs in determining \emph{steady} 
solutions in large physical domains.  As developed by Mielke to 
explore Saint-Venant's principle in elasticity 
\cite{Mielke88a,Mielke88b,Mielke88c}, the spatial dimension is treated 
as a ``time'' variable and centre manifold theory 
applied.\index{spatial evolution} The analysis elucidates spatial 
structures in the problem, such as the \idx{Ginzburg-Landau 
equation} for nearly periodic solutions.  Applications of this idea 
have been to: internal waves of a stratified fluid in a 
two-dimensional channel and capillary surface waves \cite{93f:58080}; 
Poiseuille flow between parallel plates \cite{95h:76045}; hydrodynamic 
stability in an infinitely long cylinder by Iooss \etal{} 
\cite{92b:35027,92b:35026}; the dynamics of Taylor-Couette flow also 
by Iooss \& Adelmeyer \cite{Iooss92}; and bifurcations and other 
behaviour near spatially periodic structures rather than spatially 
constant \cite{93f:76048,95f:35116}.  This trick is also employed, see 
\S\ref{Sbound}, to determine boundary conditions for models of 
spatio-temporal dynamics.

\end{itemize}

In the last of the above mentioned applications there is no time 
dependence, no dynamics!  This is a disappointing limitation.  
However, by stretching the centre manifold techniques and concepts 
past the rigorous extent of current theory, we may form models of 
spatio-temporal dynamics.  In problems with large-spatial extent, as 
discussed in the next three subsections, the resulting centre manifold 
will be of ``infinite dimension''.  Nonetheless, the model will be 
significantly simpler in one or more characteristics than the 
``infinite dimensional'' \toe{} from which it is derived---such models 
are useful.

\subsection{Slow space variations---dispersion in a channel}
\label{SSdisp}

\begin{figure}[tbp]
	\centerline{{\setlength{\unitlength}{0.92pt}
\begin{picture}(401,95)
\thinlines    \put(310,36){$c(x,y,t)$}
              \put(297,50){concentration}
              \put(202,39){diffusion}
              \put(195,52){cross-stream}
              \put(188,31){\begin{picture}(4,37)
\thinlines        \put(2,17){\vector(0,1){20}}
                  \put(2,25){\vector(0,-1){25}}
                  \end{picture}}
\thinlines    \put(103,37){$u(y)$}
              \put(92,50){advection}
              \put(65,45){\vector(1,0){20}}
              \put(65,52){\vector(1,0){20}}
              \put(65,38){\vector(1,0){19}}
              \put(65,59){\vector(1,0){19}}
              \put(65,30){\vector(1,0){16}}
              \put(65,65){\vector(1,0){16}}
              \put(65,22){\vector(1,0){12}}
              \put(65,71){\vector(1,0){12}}
              \put(65,16){\vector(1,0){6}}
              \put(65,77){\vector(1,0){6}}
              \put(17,68){$y$}
              \put(32,36){$x$}
              \put(12,48){\vector(0,1){24}}
              \put(12,48){\vector(1,0){25}}
\thicklines   \put(11,12){\line(1,0){380}}
              \put(12,83){\line(1,0){379}}
\end{picture}}}
	\caption{schematic diagram of the dispersion of a contaminant with 
	concentration field $c(x,y,t)$ in a channel with advection $u(y)$ 
	and cross-stream diffusion.}
	\protect\label{Fchan}
\end{figure}
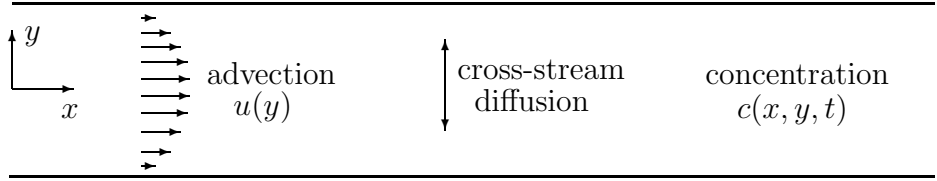

In a long, thin channel or pipe, as shown schematically in 
Figure~\ref{Fchan}, the dominant process of \idx{dispersion} of a 
contaminant is diffusion across the thin channel.
This leads to decay of all modes except the constant 
mode (across the channel).  But at different stations the constant 
will typically be different, and so the cross-sectional mean 
concentration $C$ will depend upon downstream distance $x$.  The 
shearing then causes these variations in $x$ to move and interact 
leading to the \idx{Taylor model} \cite{Taylor53,Taylor54}
\begin{equation}
	\D tC=-U\D xC +D\DD xC+\cdots \,,
	\label{Echantay}
\end{equation}
where $U$ is simply the mean downstream velocity, but the effective 
diffusivity $D$ is a nontrivial function of the velocity field.  The 
Taylor model predicts the long-term concentration is in the shape of a 
moving and spreading Gaussian.  This model is derived rigorously via 
centre manifold theory using an interesting trick introduced in 
\cite{Roberts88a}.

The simplest example is the physical problem shown in 
Figure~\ref{Fchan} which has the non-dimensional governing equation
\begin{equation}
	\D tc=-u(y)\D xc+\delta\DD xc+\DD yc\,,
	\label{Echan}
\end{equation}
where, for definiteness, the advecting velocity may be 
$u(y)=3(1-y^2)/2$ in a channel $-1<y<1$, and where $\delta$, the 
inverse square of a Peclet number, is typically very small.  Now take 
the \idx{Fourier transform} of~(\ref{Echan}), so that 
$c=\int_{-\infty}^\infty\hat ce^{ikx}dx$, to obtain the Fourier space 
equation
\begin{equation}
	\D t{\hat c}=-iku(y)\hat c-\delta k^2\hat c+\DD y{\hat c}\,.
	\label{Echanf}
\end{equation}
Then utilise the trick introduced in \S\ref{SSextend} to unfold 
bifurcations, but here to introduce the approximation of a 
\emidx{large-scale, slowly-varying} spatial dependence.  The trick is 
to adjoin to~(\ref{Echanf}) the trivial equation that the wavenumber 
$k$ is constant:
\begin{equation}
	\D tk=0\,.
	\label{Echank}
\end{equation}
Adjoining this trivial equation focuses the centre manifold analysis 
upon the small wavenumber, large-scale dynamics in the channel.  The 
advection term $-iku(y)\hat c$ is then ``nonlinear'' in $k$ and $\hat 
c$.  Linearly (in this new sense) cross-channel diffusion causes 
concentrations $\hat c$ to decay exponentially quickly to a constant 
with respect to $y$: $\hat c\approx \hat C$.  Centre manifold theory 
applied to~(\ref{Echanf}--\ref{Echank}) then derives, as a matter of 
course, that the Fourier transform of the cross-channel average, $\hat 
C$, will in the long-term evolve according to
\begin{equation}
	\D t{\hat C}=-ik\hat C-\left(\delta+\frac{2}{105}\right)k^2\hat C+\Ord{k^3}\,.
	\label{Echanmod}
\end{equation}
Taking the inverse Fourier transform leads to the Taylor 
model~(\ref{Echantay}), with here the particular coefficients $U=1$ 
and $D=\delta+2/105$.

Observe that the \idx{Taylor model}~(\ref{Echantay}), while certainly 
simpler than the \toe~(\ref{Echan}), is still of ``infinite'' 
dimension\index{infinite dimension}---there is an infinite freedom in 
$C(x)$ that is governed by the model.  In such ``infinite'' 
dimensional centre manifolds there is a much richer field of possible 
and applicable models to investigate than in centre manifolds of just 
a few dimensions.

It is straightforward for the centre manifold analysis to compute 
\idx{higher order models} of the dispersion, involving terms such as 
$C_{xxx}$ and $C_{xxxx}$ for example, that show the evolution of the 
\idx{skewness} and \idx{kurtosis} of the concentration distribution.  Indeed, 
Mercer \& I have extended the analysis of the dispersion in a channel 
\cite{Mercer90} and pipe \cite{Mercer94a} to \idx{very high order} and 
shown, using a generalisation of the \idx{Domb-Sykes plot} 
\cite[Appendix]{Mercer90}, that these models actually converge for 
non-dimensional downstream wavenumber $|k|$ less than roughly 10, 
depending upon the particular problem in hand (similar bounds also 
have been computed for beam theory, see \S\ref{Sslow} and 
\cite{Roberts93}).  Thus here we are assured that the model for 
dispersion resolves spatial\index{spatial resolution} details down to 
roughly $2\pi/10$ times the downstream advection distance in a 
cross-stream diffusion time.  This quantitative limit on the 
resolution is some 10 times better than that estimated by Taylor.  The 
Relevance Theorem also assures us that the model will resolve temporal 
\index{temporal resolution} dynamics longer than a cross-stream 
diffusion time, the time-scale of approach to the centre manifold.  
Both of these limits to the resolution of the model can only be 
improved by retaining more dynamic modes in the model, as was done by 
Watt \& I \cite{Watt94b} in investigating an invariant manifold model 
based upon the two or three gravest modes in the channel.  That we can 
sometimes find these \emph{quantitative} bounds to the domain of the 
validity of a low-dimensional model is part of the power of centre 
manifold theory.

The above analysis in Fourier space is more-or-less rigorous.  
However, upon inspecting the details of the algebra it is apparent 
that the wavenumber, in the combination $ik$, just acts as a place 
holder for the spatial derivative $\partial/\partial x$.  The 
adjoining of the trivial dynamical equation~(\ref{Echank}), that 
$\partial k/\partial t=0$, just focusses attention upon \idx{small 
wavenumber}.  Precisely the same effect is achieved in physical 
variables simply by treating $\partial_x=\partial/\partial x$ as 
``small.''  A rationale for doing this is outlined in 
\cite{Roberts88a} based upon the local dynamics in any reach of the 
domain, but rigorous it is not because the operator $\partial_x$ is 
unbounded and the extant theory \cite{Gallay93} cannot be applied 
directly.  However, an analysis based upon this idea has the advantage 
that one deals in physical variables, rather than Fourier transformed 
quantities, and it easily generalises to more difficult problems as 
seen in the next two subsections.  For example, one may analyse 
\cite{Mercer90,Mercer94a} the effect upon the Taylor model of spatial 
and temporal variations in the flow-rate, width of the channel, and 
cross-stream diffusion.  Interestingly, as first pointed out by Smith 
\cite{Smith83b}, the results may be recast in terms of a \idx{memory} of the 
conditions upstream or at earlier times.

The centre manifold analysis of dispersion in a pipe has also been 
extended by Balakotaiah \& Chang \cite{Balakotaiah92} to the dynamics 
in a \idx{chromatograph} chemical reactor where chemical reactions occur 
either in the flow or with the walls of the pipe.

\subsection{Cross-sectional averaging is unsound---thin film flows}
\label{SSfilm}

\begin{figure}
	\begin{center}
{\tt    \setlength{\unitlength}{0.92pt}
\begin{picture}(370,141)
\thinlines    \put(20,19){\vector(0,1){107}}
              \put(13,28){\vector(1,0){63}}
\thicklines   \put(82,28){\line(1,0){278}}
\thinlines    \put(73,14){$x$}
              \put(10,122){$y$}
              \put(159,12){solid, $u=v=0$}
              \put(24,87){\line(1,0){34}}
              \put(58,87){\line(6,1){47}}
              \put(105,95){\line(3,1){38}}
              \put(143,108){\line(6,1){43}}
              \put(186,115){\line(1,0){35}}
              \put(221,115){\line(6,-1){48}}
              \put(269,107){\line(1,0){38}}
              \put(307,107){\line(6,-1){49}}
              \put(49,99){$y=\eta(x,t)$}
              \put(138,119){atmospheric pressure}
              \put(170,77){Navier-Stokes equation}
              \put(341,59){$h$}
              \put(205,56){$p(x,y,t)$}
              \put(115,56){$u(x,y,t)$}
              \put(337,64){\vector(0,1){33}}
              \put(337,65){\vector(0,-1){36}}
              \put(89,88){\vector(1,0){25}}
              \put(89,75){\vector(1,0){22}}
              \put(89,62){\vector(1,0){19}}
              \put(89,49){\vector(1,0){13}}
              \put(89,36){\vector(1,0){6}}
\end{picture}}
	\end{center}
	\caption{schematic diagram of a thin fluid film flowing down a solid bed.}
	\protect\label{Ffilm}
\end{figure}
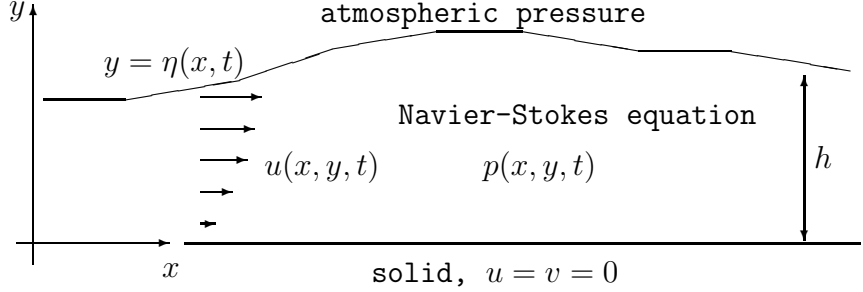

The dynamics of \idx{thin films} of fluid are important in many 
industrial, environmental and biological processes.  An approximation 
of such a thin viscous fluid flow, as shown schematically in 
Figure~\ref{Ffilm}, with slow spatial variations leads to a 
Ku\-ra\-mo\-to-Si\-va\-shin\-sky type of 
equation.\index{Kuramoto-Sivashinsky} If \idx{surface tension} 
$\sigma$ is the only driving force then the simplest long-wave model 
is
\begin{equation}
	\D t\eta\approx -\frac{\sigma}{3\mu}\D x{\ }\left(\eta^3\Dn 
	x3\eta\right)\,,
	\label{Efilm1}
\end{equation}
where $\mu$ is the fluid viscosity and $\eta(x,t)$ is the film 
thickness above a flat substrate.

To analyse the \idx{Navier-Stokes equations}, the \toe{}, to derive such a 
model \cite{Roy96}, recognise that across the thin fluid film, 
viscosity acts quickly to damp almost all cross-film structure.  The 
distinction between the longitudinal and cross fluid dynamics is 
exactly analogous to that of dispersion in a pipe; we proceed with 
a similar analysis.  However, the equations are very different (see 
Figure~\ref{Ffilm}).  In particular, this problem has many 
nonlinearities: not only is the advection in the Navier-Stokes 
equation described by a nonlinear term, but also the thickness of the 
fluid film is to be found as part of the solution and its unknown 
location is another source of nonlinearity.  Here, the only practical 
course of analysis is to deal with physical variables and treat 
$\partial_x$ as a ``small'' parameter which is negligible to the 
leading ``linear'' approximation (though recall that H\u{a}r\u{a}gus 
\cite{Haragus95b} otherwise justifies the Korteweg-de Vries equation for the 
\emph{inviscid}, long-wave hydrodynamics).

Assuming no longitudinal variations at all, a linear analysis of the 
equations shows that there is one critical mode in the cross-fluid 
dynamics, all others decay exponentially due to viscosity.  This 
critical mode is associated with conservation of the fluid and 
consequently it is natural to express the low-dimensional model in 
terms of the film thickness $\eta(x,t)$.  An interesting aspect is the 
fact that although this is a nonlinear problem, \idx{conservation of fluid} 
applies no matter how thick the fluid layer, and the state of no flow 
is an equilibrium for all constant $\eta$.  Thus the analysis is valid 
for arbitrarily large variations in the thickness of the film, just so 
long as the variations are sufficiently slow in space and time.  A 
relatively simple toy example of this is also discussed in 
\cite{Roberts88a}.  These are examples, as mentioned in 
\S\S\S\ref{SSSapprox}, of a centre manifold analysis based upon a 
\idx{manifold of equilibria} rather than being anchored to just one fixed 
point.

Treating $\partial_x$ as a small perturbing operator, the centre 
manifold formalism follows to straightforwardly derive models such 
as~(\ref{Efilm1}).  Of course, \idx{higher order models} may be 
straightforwardly constructed if necessary \cite{Roberts96a}, viz
\begin{eqnarray*}
	\D t\eta &=&-\frac{\sigma}{\mu}\D x{\ }\left[\frac{1}{3}\eta^3\eta_{xxx}
	+\frac{3}{5}\eta^5\eta_{xxxxx} +3\eta^4\eta_x\eta_{xxxx} 
	\right.\\&&\qquad\left.
	 +\eta^4\eta_{xx}\eta_{xxx} +\frac{11}{6}\eta^3\eta_{x}^2\eta_{xxx} 
	 -\eta^3\eta_x\eta_{xx}^2 \right]
	 +\Ord{\partial_x^7}\,.
\end{eqnarray*}
However, much more interesting is to model the 2D spread of a 3D 
fluid over a \idx{curved substrate} \cite{Roy96}.  Provided that, as in 
dispersion in a varying width pipe or channel 
\cite{Mercer90,Mercer94a}, the length-scale of curvature variations is 
large compared to the film thickness, a similar analysis leads to the model
\begin{equation}
		\D t\zeta\approx -\frac{\sigma}{3\mu}\nabla\cdot
		\left[ \eta^{2}\zeta\nabla\tilde\kappa
	-\frac{1}{2}\eta^{4}(\kappa {\vec I}-\vec K)\cdot\nabla\kappa \right]\,,
	\label{Ecurvfilm}
\end{equation}
where: $\zeta= \eta-\frac{1}{2}\kappa\eta^2 + \frac{1}{3} k_{1} k_{2} 
\eta^{3}$ is proportional to the amount of fluid locally ``above'' the 
substrate; $\tilde\kappa$ is the approximate mean curvature of the 
free-surface; $\vec K$ is the curvature tensor of the substrate; $k_{1}$, 
$k_{2}$, and $\kappa = k_{1}+k_{2}$ are the principal curvatures and 
the mean curvature of the substrate, respectively; and the 
$\nabla$-operator is expressed in a coordinate system of the curving 
substrate.  The effects of gravity and fluid inertia also may be included 
systematically \cite[\S4]{Roy96} to produce a more extensive model.  

\emph{One significant limitation in the use of the whole family of 
models discussed above is that they only resolve\index{temporal 
resolution} dynamics significantly slower than the time-scale of 
viscous decay, $T\approx 0.4\eta^2/\nu$, of the gravest shear mode, 
$u_1 \propto \sin (y\pi/2\eta)$.} For thicker fluid films, interesting 
dynamics occurs on about this time-scale.  Another class of models is 
needed in such a situation.  Traditionally these are obtained by 
integrating or averaging the 
horizontal momentum equation over the depth of the fluid to obtain an 
evolution equation for the mean horizontal velocity $\bar u(x,t)$ 
\cite[e.g.]{Prokopiou91b,Chang94}.

\begin{figure}[tbp]
	\centerline{\includegraphics{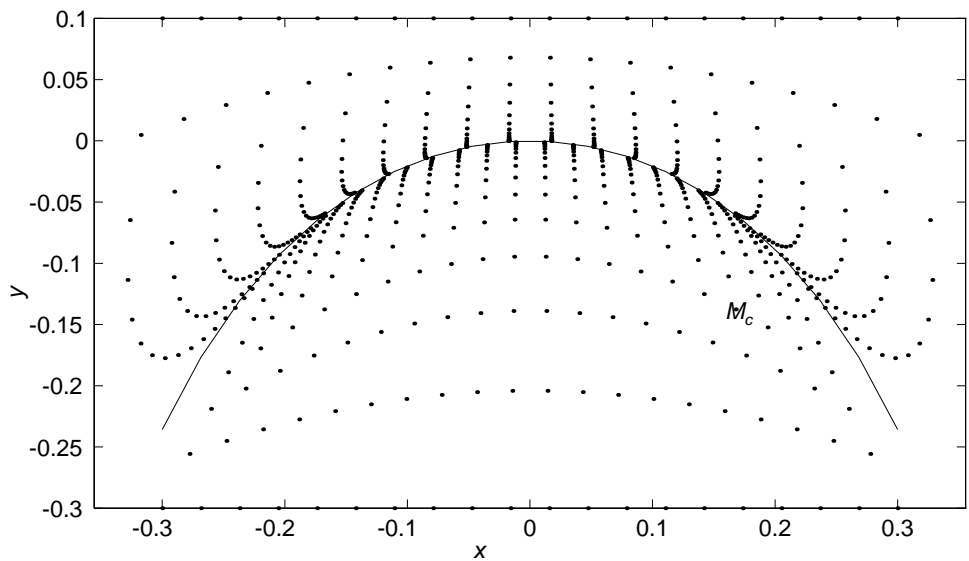}}
	\caption{trajectories of the dynamical system 
	(\protect\ref{Eegave}), plotted as dots $\Delta t=0.4$ apart, 
	showing that although trajectories point away from the origin on 
	$y=0$, nonetheless the nonlinear dynamics on the centre manifold, 
	$\cM_c$, shows that the origin is stable.}
	\protect\label{Fegave}
\end{figure}

Unfortunately, such simple modelling methods do not just get the 
coefficients wrong, they can fail dramatically.  Abstractly, 
\idx{cross-sectional averaging} is a projection of the dynamical 
equations onto some linear subspace of the state space, namely the 
space of functions which are constant over cross-sections.  Typically, 
linear considerations suggest that this would be a useful procedure.  
But consider the simple example \cite[p153]{Kuznetsov95}
\begin{equation}
	\dot x=xy+x^3\,,
	\quad\mbox{and}\quad
	\dot y=-y-2x^2\,,
	\label{Eegave}
\end{equation}
whose trajectories are plotted in Figure~\ref{Fegave}.
Linearly, $\dot x=0$ and $\dot y=-y$ and so solutions 
exponentially quickly approach the subspace $y=0$.  Thus, projecting 
the dynamics onto $y=0$ to give $\dot x=x^3$ may be expected to give a 
reasonable model of the long-term dynamics of the system from which 
one would deduce that $x=0$ is an unstable fixed point.  But this is 
not so.  The centre manifold is $y=-2x^2-4x^4+\Ord{x^6}$ whence we 
deduce the correct model of the long-term dynamics to be $\dot 
x=-x^3+\Ord{x^4}$ and hence the origin is actually stable, as seen in 
Figure~\ref{Fegave}.  \emph{Cross-sectionally averaging dynamical 
equations is unsound as a modelling paradigm.}

Returning to the dynamics of a fluid film, observe that in the 
horizontal shear modes, $u_n\propto\sin(ny\pi/2\eta)$ there is a 
relatively large \emidx{spectral gap} between the gravest mode $u_1$ 
and the next mode $u_2$; the eigenvalues are 
$\lambda_1=-2.5\nu/\eta^2$ and $\lambda_2=-22.2\nu/\eta^2$ 
respectively.  Surely we should be able to gather the $u_1$ mode into 
the centre manifold.

The trick of \S\S\ref{SSextend} allows us to apply centre manifold 
techniques to obtain a coupled model for the evolution of $\bar u$ and 
$\eta$ which resolves transients on a shorter time-scale.  
Manipulating the horizontal momentum equation \cite{Roberts94c} to
\begin{equation}
	\D tu=-u\D xu-v\D yu-\D 
	xp+\nu\nabla^2u\quad+(1-\gamma)\nu\left(\frac{\pi}{2\eta}\right)^2u\,,
	\label{Eumom}
\end{equation}
and adjoining $\dot\gamma=0$ changes the spectrum.\index{spectrum 
manipulation} When $\gamma=0$, the introduced artificial forcing makes 
$u_1$ a critical mode, along with $\eta$, and hence there exists a 
centre manifold parameterised by $\avu$, measuring the amplitude of 
$u_1$, the film thickness $\eta$, and the artificial parameter 
$\gamma$.  Setting $\gamma=1$ recovers the original \pde{}s and so 
pursuing the analysis and subsequently setting $\gamma=1$ leads to an 
approximate model for the original dynamics.  The model is found 
\cite{Roberts94c} to be
\begin{eqnarray}
\D t\eta&=&-\D x{\ }\left(\avu\eta\right)\,,\\
\D t\avu&\sim& 0.8238 \left(- g\eta _x + \sigma \eta _{xxx}\right)
	-1.504  \avu\avu_x
	-2.467 \frac\nu{\eta^2} \avu
	-0.1516  \frac{\avu^{2}}{\eta} \eta _x\,,
	\label{Eavudt}
\end{eqnarray}
as in depth averaged equations but quantitatively different.  The 
adaptation~(\ref{Eumom}) of the horizontal momentum equation is not 
unique.  Exactly the same eventual model~(\ref{Eavudt}) is obtained  
\cite{Roberts96b} by appropriate manipulation of the tangential stress 
boundary condition at the free-surface.

Similar ideas are being employed to analyse the dynamics of a 
\idx{turbulent river} or flood.  Initial work by Mei \& I is reported in 
\cite{Mei94} where we took the $k$-$\epsilon$ model of turbulent flow 
as the \toe, and perturbed some critical coefficients so that a centre 
manifold inspired model could be constructed based upon the water 
depth $\eta$ and the depth averages of the horizontal velocity $\bar 
u$, the turbulent energy $\bar k$, and the turbulent dissipation $\bar 
\epsilon$.  Further analysis on this interesting model is in 
progress; it promises a sophisticated and reliable model of flood, 
river and estuarine dynamics.

However, we need more powerful theory on infinite dimensional centre 
manifolds to provide rigorous support for these sorts of interesting 
slowly-varying, long-wave models.  I tell my graduate students that as 
far as rigorous theory is concerned: for simple bifurcations we are on 
solid ground, for shear dispersion we may well be on thin ice, but in 
the application to thin film flows we are walking on water!

\subsection{Bands of critical modes---convection}
\label{SSband}

In a fluid layer heated from below, diffusion may damp all motion.  If 
the heating is large enough then warm light fluid rises and cool heavy 
fluid falls.  If the top and bottom boundaries are insulating, the 
case of \idx{fixed heat flux}, as elaborated by Proctor \etal{} 
\cite{Chapman80a,Chapman80b,Proctor81a,Depassier82,Roberts85c}, the 
\idx{convection} occurs on a large horizontal length-scale, everywhere 
\idx{small wavenumber}, and so it may be modelled by the approach of 
the previous subsection.

But in the usual case of \idx{fixed temperature} top and bottom boundary 
conditions, the fluid flow occurs on a horizontal length-scale of the 
same size as the height of the fluid layer.  Near the critical 
temperature difference (measured by the \idx{Rayleigh number} $R$), and 
using the trick of \S\S\ref{SSextend} again, a centre manifold may be 
found consisting primarily of a superposition of ``rolls'', say
\begin{equation}
	u\approx A\ell e^{ikx}\cos(\ell z)+\mbox{c.c.}\,,
	\quad\mbox{and}\quad
	v\approx Aik e^{ikx}\sin(\ell z)+\mbox{c.c.}\,,
	\label{Erolls}
\end{equation}
for some horizontal wavenumber $k$ and some vertical wavenumber 
$\ell$, where $A$ is the \idx{complex amplitude} of the rolls and 
``c.c.''  denotes the complex conjugate of the appearing terms.  The 
centre manifold analysis \cite[e.g.]{94c:76040,91g:76038} produces a 
\emidx{Landau equation} for the amplitude such as
\begin{equation}
	\frac{dA}{dt}=\left(R-R_c\right)A-\alpha|A|^2A+\Ord{A^4+\epsilon^2}\,.
	\label{Elandau}
\end{equation}
However, there are serious problems with such an application in the 
usual interesting case of convection with large horizontal extent.  
These problems are endemic to a wide range of \idx{pattern evolution} 
problems as discussed in the review by Newell \etal{} \cite{Newell93}.
\begin{figure}[tbp]
	\centerline{\includegraphics{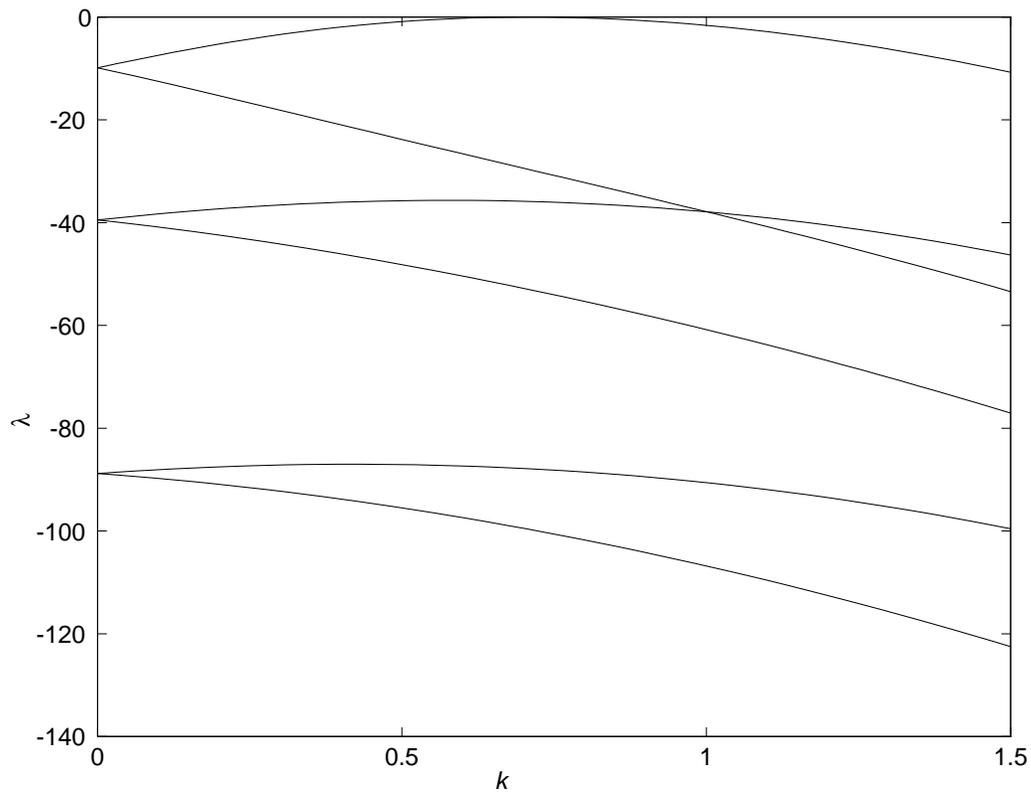}}
	\caption{eigenvalues of all modes in convection exactly at 
	critical Rayleigh number versus the continuum of horizontal 
	wavenumbers, $k$, possible in a layer of large extent.  The 
	different branches denote different modes in the vertical.  This 
	particular graph is for stress free boundaries and a Prandtl 
	number of~1.}
	\protect\label{Frbeig}
\end{figure}%
The critical spectrum typically looks like that plotted in 
Figure~\ref{Frbeig}.  Almost all modes decay rapidly and so we should 
be able to form a low-dimensional model.  However, there are a 
continuum of modes with wavenumbers $k$ close to the critical (here 
$k_c=1/\sqrt 2$), and for $R>R_c$ a finite band of these modes 
become weakly unstable.  As in the previous subsections, there should 
be some way to use the centre manifold techniques and concepts to 
justify creating a low-dimensional model such as~(\ref{Elandau}).

One avenue of rigorous application of centre manifold theory is 
mentioned in the start of this section.  If we choose to seek just 
\emph{steady} convective patterns, then we may treat the spatial 
variable $x$ as a ``time-like'' variable and construct 
\cite{93f:76048,95f:35116} \ode{}s that model the spatial 
``oscillations'' of the roll type structure in convection.  
\index{spatial evolution} Unfortunately, although useful for some 
purposes this approach seems rather limited: it does not address the 
dynamics, just the equilibria; and there seems no straightforward 
generalisation to the analysis of the 2D planform adopted by a 3D 
fluid.

At this stage, one useful approach to model the dynamics is to involve 
a mixture of centre manifold techniques and ideas from the method of 
\idx{multiple scales}.  There is no real rigor as yet in general.  Use two space 
scales: the short space scale $x$ to resolve the small-scale structure 
of the rolls; and a large space scale, $X$ say, to resolve the 
large scale modulation of the rolls implied by the band of unstable 
wavenumbers.  Then horizontal derivatives, $\partial_x$ in the 
Navier-Stokes equations, the \toe{}, become $\partial_x+\partial_X$.  
We then treat $\partial_X$ as ``small'' just as in the previous 
subsections.  The centre manifold algebra then proceeds 
straightforwardly to deduce a 
\emph{Ginzburg-Landau}\index{Ginzburg-Landau equation} model
\begin{equation}
	\D tA \approx \left(R-R_c\right)A-\alpha |A|^2 A+\beta\DD XA+\cdots\,,
	\label{Eginzlan}
\end{equation}
for the amplitude of the rolls.  Such Ginzburg-Landau models feature 
prominently in investigations of pattern evolution, see the review by 
Cross \& Hohenberg \cite[\S{}III.C.2.d]{Cross93}.  Recently Eckhaus 
\cite{Eckhaus93} has proved that the Ginzburg-Landau equation is 
indeed relevant to one-dimensional pattern evolution.  This centre 
manifold approach is superior to that of multiple scales alone, not 
only because of the better geometric viewpoint of centre manifold 
theory, but also because higher-order corrections\index{higher order 
models} may be computed without introducing even further space and 
time scales \cite[Eqn.(2.3),e.g.]{Fujimura}---just the one extra space 
scale is sufficient.  Again, however, more theory is needed to support 
the algebraic formalism in the application of this idea to general 
problems and their models.

The derivation of models of the spatially 1D evolution of patterns, 
such as the Ginzburg-Landau model~(\ref{Eginzlan}) for 2D fluid 
convection, is reasonable.  However, the low-dimensional modelling of 
patterns in two spatial dimensions, as in 3D fluid convection, is 
considerably more subtle \cite{Newell93}.  To investigate 2D 
\idx{pattern evolution}, it is natural to look at the modes and their 
interactions in 2D Fourier space with wavenumber $\vec k$.  The 
difficulty in 2D pattern evolution, and I use convection as a specific 
example, seems to stem from the fact that the critical modes do not 
come from a localised band of wavenumbers, but from all the way around 
an annulus that extends a finite size in wavenumber space, $|\vec 
k|\approx k_c$.  The nonlinear interactions among such an annulus of 
modes are vastly richer than those among the small lump of critical 
modes in a 1D pattern evolution.  No really satisfactory modelling 
procedure has yet been developed, at least not to my taste.

A satisfactory model should only resolve the slow evolution of the 
near critical modes.  Failing that, one possibility is to carry some 
dynamical ``\idx{dead wood}'' in the model by also resolving modes, 
maybe unphysical modes, which are exponentially decaying.  This 
concept is developed in~\cite{Roberts92b} where it was linked to the 
geometric idea of \emph{\idx{embedding} a centre manifold} and the 
evolution thereon within the dynamics of a higher dimensional 
dynamical system, but nonetheless of lower dimension than the original 
\toe{}.  There I show that \emidx{adiabatic iteration}, namely the 
repeated application of \idx{adiabatic 
elimination}~\cite{Haken83,vanKampen85,Titi90}, is an effective 
algebraic procedure to do this embedding.  An open question is: are 
there other, ``better'' embedding procedures?  In problems of the 
convection type, the adiabatic iteration embedding procedure leads to 
a \emph{generalised \idx{Swift-Hohenberg equation}}~\cite{Swift77},
\begin{equation}
	\frac{\partial a}{\partial t}\approx(R-R_c) a
	-\left(k_c^2+\nabla^2\right)^2a
	-a\cG\star a^2\,,
	\label{Eshreq}
\end{equation}
where $\cG\star$ is some particular radially symmetric convolution.  
The field $a(\vec x,t)$ has the same critical annulus of modes, $|\vec 
k|\approx k_c$, in which the linear dynamics are that of the \toe{}.  
Modes away from the critical annulus are irrelevant as they decay 
rapidly, they are the ``dead wood.''  The correct interaction among 
the critical modes is obtained through the correct determination of 
the \idx{nonlocal nonlinearity} in the convolution.  Such a model is a 
remarkably accurate predictor of the pattern 
evolution~\cite{Roberts92c} in a toy convection problem.  
Interestingly, people who invoke \idx{symmetry} arguments to derive 
the Swift-Hohenberg equation generally fail to acknowledge the 
possibility of such a nonlocal nonlinearity even though it is indeed 
permitted \cite[p30,e.g.]{Greenside84}.

Such spatio-temporal modelling of pattern evolution is perhaps a 
precursor to the modelling of \idx{turbulence} with its variations 
across a wide range of spatial and temporal scales.  One avenue I 
would love to find time to explore is to express turbulent fluid flow 
as a field of interacting \idx{wavelets} and then embed the dynamics 
within some economical description.  The idea is that the wavelets 
will resolve the wide range of space-time scales, and one of the 
techniques described herein will show how to model the interaction.

\section{Competing small effects should be independent}
\label{Ssmall}

It is characteristic of many interesting physical problems that there 
are several ``small'' parameters.\index{small parameters} In the 
Ginzburg-Landau equation~(\ref{Eginzlan}) for the pattern evolution of 
convection near onset there is $R-R_c$, the amplitude $A$, the spatial 
derivative $\partial_X$, and potentially the along-roll spatial 
derivative $\partial_y$.  In the flow of thin fluid films over a 
curved substrate, (\ref{Ecurvfilm}), there is the curvature of the 
substrate, the gradients of the free-surface, and potentially the 
Reynolds number, gravitational forcing, surface contamination, etc.  
The centre manifold approach enables a rational treatment of many and 
varied small effects, and then allows any consistent truncation when 
the model is applied.

\subsection{Traditional scaling is restrictive}

In the construction of dynamical models it is traditional to scale 
\emph{a priori} all such small effects together in terms of a 
\emph{single} small parameter, say $\epsilon$.  For example, to obtain 
the \idx{Ginzburg-Landau equation} in \idx{convection}: 
$R-R_c=\epsilon^2$, $A=\Ord{\epsilon}$, $\partial_X=\Ord{\epsilon}$, 
and $\partial_y=\Ord{\sqrt\epsilon}$ \cite[eqn.(4.6)]{Cross93}.  For 
thin film\index{thin films} flows, the substrate curvature 
$\kappa=\Ord{\epsilon^2}$, $\partial_x=\Ord{\epsilon}$ 
\cite{Schwartz95}, and, if not neglected entirely, gravity 
$g=\Ord{\epsilon^2}$ unless the substrate is nearly all horizontal 
when $g=\Ord{\epsilon}$.  This is done so that all the interesting 
dynamical effects occur at the one order in $\epsilon$, namely the 
leading order.  But, to give just one example, in the spatially 
extended system of thin film flow there may be regions where 
surface tension through substrate curvature is the dominant forcing, 
and other regions, where the substrate curvature is constant, in which 
gravity is the dominant forcing, and further the substrate could be 
nearly horizontal in some places and nearly vertical in others.  
Demanding that all interesting effects be scaled to appear at leading 
order is too restrictive.  It is a ``straight-jacket'' that 
traditional techniques force upon us at the outset of the modelling.

The situation may be even worse.  Sometimes the leading order model is 
structurally unstable, as in the \idx{Taylor model} of 
dispersion~(\ref{Echantay}) which to leading order in $\partial_x$ is 
just the advection equation $\partial_tC=-U\partial_xC$.  Higher order 
corrections, not appearing at leading order, are necessary to obtain a 
reasonable, structurally stable\index{structural stability} model.  
For simple problems, such as shear dispersion 
\cite[Eqn.(17)]{Taylor53} and propagating waves \cite[p9]{Peregrine85} 
or convection rolls, the trick of transforming the \toe{} to a 
reference frame moving with an appropriate velocity, $U$ in the case 
of shear dispersion and the group velocity in the case of waves, 
causes the low-order advection to disappear and the leading order in 
the model is then the higher-order terms needed for structural 
stability.  However, in experiments one may have interacting waves or 
rolls travelling with different speeds or even in completely different 
directions.  A moving reference frame cannot assist the analysis for 
these problems.  \emph{The traditional scaling paradigm, of using one 
small parameter to scale all others and then requiring all effects to 
appear at leading-order, is fundamentally flawed.}

\subsection{A consistent flexibility}

In contrast, centre manifold theory asserts, through 
Theorem~\ref{Tcmapprox}, that one can approximate the shape of the 
centre manifold and the evolution thereon to any order in amplitude 
$s$.  Competing small effects may appear at any order in the analysis, 
they need not just arise at leading order; after all the \idx{Taylor 
series} for $\exp(x)$ is $1+x+x^2/2+\cdots$ to any convenient order.  
It is then consistent to include all terms up to a specific order in 
the model.

However, with parameters one may be considerably more flexible.  For 
example, Theorem~\ref{Tcmparam} shows that the model describing the 
\idx{pitchfork bifurcation} in
\begin{equation}
	\dot x=\epsilon x-xy\,,
	\quad\mbox{and}\quad
	\dot y=-y+x^2\,,
	\label{Eegalso}
\end{equation}
may be written variously as
\begin{eqnarray*}
\dot x & = & \epsilon x-x^3+\Ord{x^5+\epsilon^{5/2}}  \\
	 & = & \epsilon x-(1-2\epsilon)x^3+\Ord{x^5,\epsilon^2} \\
	 & = & \epsilon x-(1-2\epsilon)x^3-2x^5+\Ord{x^7+\epsilon^{7/2}}\,.
\end{eqnarray*}
These are all \emidx{consistent truncations} of the following multivariate 
asymptotic expansion\index{multivariate expansion}\index{higher order 
models}
\begin{eqnarray*}
	\dot x & = & x\epsilon  \\
	 &  & -x^3(1-2\epsilon+4\epsilon^2-8\epsilon^3+16\epsilon^4)  \\
	 &  & -x^5(2-16\epsilon+88\epsilon^2-416\epsilon^3+1824\epsilon^4) \\
	 &  & +\cdots\,.
\end{eqnarray*}
As noticed by Coullet \& Spiegel \cite{Coullet83}, this feature is 
very important.  The centre manifold approach allows you to compute as 
many orders as you like, different orders in the amplitudes and the 
parameters (even different for the different parameters), and then 
\emph{when you come to \emph{use} the model, you may choose any 
consistent truncation that is appropriate for the particular 
realisation you wish to investigate.}

Such freedom is immensely valuable when there is any more than a 
couple of physical parameters in the problem.

\section{The slow manifold is central}
\label{Sslow}
\index{slow manifold}

In a purely elastic body, elastic waves ``ring'' perpetually within 
the body.  If these high frequency vibrations are ignored, then what 
is left is the relatively slow dynamics of rigid body motion.  The 
flight of a ball is an example already mentioned in \S\ref{Sintro}.  

Muncaster and Cohen \cite{Muncaster83c,Cohen88} suggested the 
construction of the low-di\-men\-sion\-al manifold of slow, \idx{rigid-body 
dynamics} by neglecting the fast modes.  The extremely simple example 
of the motion of a one-di\-men\-sion\-al elastic body is discussed in 
\cite[\S2]{Roberts93}.  In contrast to the rapid collapse to the 
centre manifold, the slow dynamics on the slow manifold form a 
low-dimensional model because they act as a ``centre'' for the fast 
oscillations of neighbouring trajectories, see Figure~\ref{Florenz} 
for example.  This principle of neglecting fast oscillations is 
completely equivalent to the \emidx{guiding centre} principle of Van 
Kampen \cite{vanKampen85}, which is frequently invoked in \idx{plasma 
physics}, see \cite{95i:76114,95a:76098,94c:76082} for some recent 
work.

\subsection{The linear basis---beam models}
\label{SSbeam}

The construction of such a slow manifold uses exactly the same 
techniques as described earlier.  It is based on the Approximation 
Theorem~\ref{Tcmapprox}, or~\ref{Tcmparam}, with the distinction that 
the \emidx{slow manifold} ($\cM_0$) has the \emidx{slow subspace} 
($\cE_0$) as its tangent space at the origin, instead of the centre 
subspace.  The slow subspace being that space spanned by the 
eigenvectors and generalised eigenvectors of the \emph{precisely zero} 
eigenvalues.  For example, consider the dynamics of a long thin 
\idx{elastic beam}.  All the vibrations are fast except for the large-scale 
flexure, torsion and displacement of the beam.  Applying the same 
ansatz of slowly-varying dependence along the beam \cite{Roberts93}, 
similar to that discussed in \S\S\ref{SSdisp} for dispersion, we may 
identify an 8-di\-men\-sion\-al slow subspace in the cross-sectional 
elastic dynamics: displacements and velocities sideways (two), 
longitudinally and rotationally.  Coupled with the assumed slow 
variations along the beam, the analysis holistically constructs 
dynamical models for the beam bending, torsion and stretching.  In 
linear elasticity of a circular beam all these models decouple 
\cite{Roberts93}.  However, for a nonlinear or non-circular beam,  
the analysis naturally couples the dynamics.  Computing the slow 
manifold to various asymptotic orders results in models equivalent to 
the range of models of classic beam theory.  \emph{The concept of a 
slow manifold is also useful in forming low-dimensional 
models of dynamics.}

You may have noted that all the specific examples of centre manifolds 
discussed in earlier sections have also been slow manifolds as they 
are based upon zero-eigenvalue modes.  However, in this section I 
specifically focus on situations where the slow dynamics occurs among 
fast oscillations.  That is, we take the spectrum of the linear 
dynamics to consist entirely of some zero eigenvalues and some purely 
imaginary eigenvalues.

\subsection{Nonlinear problems---geostrophy}
\label{SSgeost}

\begin{figure}[tbp]
	\centerline{\includegraphics{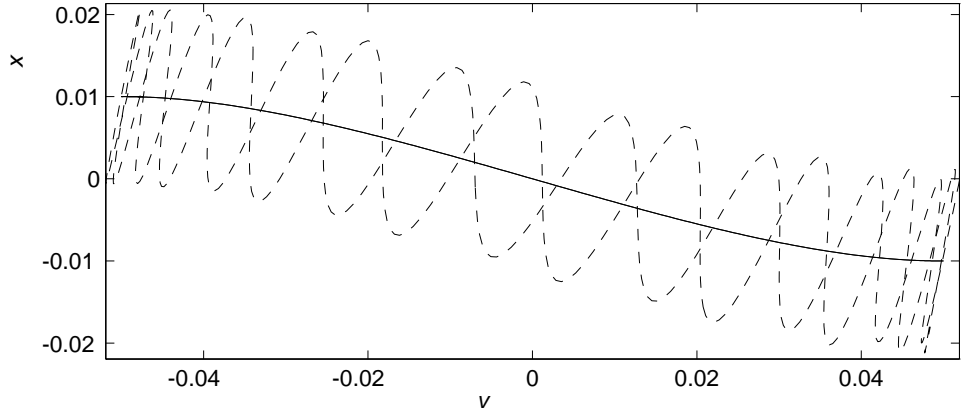}}
\caption{a comparison of the trajectories on (solid) and off (dashed) 
the slow manifold $\cM_0$ for the Lorenz 
system~(\protect\ref{Elorenz}).  Observe that the fast oscillations 
off the slow manifold reasonably track the evolution on $\cM_0$.}
	\protect\label{Florenz}
\end{figure}%
The above applications have so far been to notionally linear problems.  
The slow manifolds of nonlinear dynamics are also of interest.  This 
is shown, for example, in the importance of the concept of  
\idx{geostrophy} in atmospheric dynamics, see recent work in 
\cite{95m:76018,94d:86006}.  Lorenz \cite{Lorenz86} introduced the 
five mode dynamical system\index{Lorenz equations}
\begin{eqnarray}
\dot{u} & = & -vw+bvz\nonumber\\
\dot{v} & = & uw-buz\nonumber\\
\dot{w} & = & -uv\label{Elorenz}\\
\dot{x} & = & -z\nonumber\\
\dot{z} & = & x+buv\nonumber\,,
\end{eqnarray}
to illuminate the nonlinear slow manifold of \idx{quasi-geostrophy}.  
Observe that $x$ and $z$ oscillate quickly, while $u$, $v$ and $w$ 
evolve slowly, at least near the origin.  A couple of trajectories of 
this system are plotted in Figure~\ref{Florenz}.  The slow manifold of 
this 5-dimensional \toe{} is 3-dimensional based upon the three 0 
eigenvalues associated with $u$, $v$ and $w$ in the linear dynamics.  
Theory by Sijbrand \cite[\S6--7]{Sijbrand85} is relevant to the 
existence of a slow manifold for~(\ref{Elorenz}), namely
\begin{displaymath}
x= -buv+\Ord{s^4}, \quad z= b(u^2-v^2)w+\Ord{s^5}\,,
\end{displaymath}
where $s=|(u,v,w)|$, to some level of smoothness.  Lorenz later argued 
\cite{Lorenz87} that truly slow dynamics do not exist 
in~(\ref{Elorenz}) because if it did then there must be an infinite 
number of \idx{singularities} in the slow manifold.  However, Cox \& I 
\cite{Cox93a} have shown that the singularities are exponentially weak 
and so are negligible for small enough amplitude flow.  Recently, 
Bokhove \& Shepherd \cite{Bokhove96}, Boyd \cite{95a:76018}, Camassa 
\cite{Camassa95}, and Lorenz \cite{94a:34051} have further elucidated the 
structure and dynamics associated with the slow manifold 
of~(\ref{Elorenz}), and a connection with \idx{inertial manifolds} has been 
made by Debussche \& Temam \cite{92f:58161,93a:47073}.

The concept of a slow manifold is generally applicable to 
\idx{Hamiltonian dynamical systems} because of their typical spectrum.  
Examples are found in atmospheric dynamics \cite{Salmon85}, water 
waves \cite{Craig94,Nore96}, and plasma physics 
\cite{95i:76114,93j:78012,91k:82063,91a:78006}.

All the above issues are intriguing and deserve further study, but 
there is a further twist in such modelling: there is no 
\idx{relevance} theorem for the low-dimensional dynamics on the slow 
manifold.  There is no rigorous assurance that they do indeed model 
the \toe{}.  Instead, Cox \& I \cite{Cox93a,Cox93b} used \idx{normal 
forms}, see \S\S\ref{SSnorm}, to show that the dynamics on and off the 
slow manifold generally differ by an amount of $\Ord{r^2}$, where $r$ 
measures the amplitude of the fast oscillations, it measures the 
distance off $\cM_0$.  That is, there is some unavoidable slip between 
the model and the \toe{}.  Thus, in general, \emph{one can only expect 
a slow manifold model to be accurately predictive for a time 
$o\left(1/r^2\right)$.}

Lastly, I mention that Sijbrand \cite{Sijbrand85}, building upon work 
by Lyapunov, actually proves theorems about \emidx{sub-centre 
manifold}s.  That is, manifolds based upon the eigenspace of a pair of 
pure imaginary modes; eigenvalues that are precisely 0 are a special 
case.  His theorems are directly relevant to the existence and 
construction of \idx{nonlinear normal modes} of oscillation as 
investigated by Shaw \etal{} \cite{Shaw93,Shaw94,Shaw94b,Nayfeh94}.  
It may be that such theorems also provide the basis for a 
justification of modulation equations describing the slow space-time 
evolution of the amplitude and phase of nonlinear \idx{dispersive 
waves}.  A simple example being the \idx{nonlinear Schr\"odinger 
equation} usually derived using the method of multiple scales.  
However, a sub-centre manifold approach to the modulation and 
interaction of nonlinear waves has been elucidated in 
\cite{Roberts92}.

\section{Initial conditions are long-lasting}
\label{Sinitial}

Many low-dimensional models are used simply to explore the range of 
dynamical possibilities.  For example, in control applications one 
wants to be assured that the specified control scheme will stabilise the 
system.  However, many models are used to make definite predictions 
of later times given that the system starts from a given initial 
state---the \idx{forecast problem}.  So, suppose you know the initial 
state of the system for the \toe{}: what is the corresponding \idx{initial 
condition} for the model?

For example, if a pollutant is released into a river from a given 
site, we would wish to know what will be the \idx{dispersion} of this 
specific cloud of pollutant.  If the pollutant is released in the 
middle of a channel it will be initially carried downstream quicker 
than if it is released at the side; in the long-term the pollutant 
clouds will have different mean locations.  The initial condition used 
for the model should be able to reflect such differences in the 
release.  Appropriate initial conditions are also crucial in the 
derivation of correct boundary conditions, see \S\ref{Sbound}.

Remarkably, this issue of providing the correct initial conditions for 
a low-dimensional dynamical model has received very little attention 
in the past.  In many cases this is because attention has focussed on 
the typical dynamics inherent in a model.  However, even when 
interested in making definitive forecasts, generally people have 
simply assumed that the provision either is according to the linear 
dynamics or is simply by the evaluation of the ``amplitudes'' in the 
model.  That these assumptions are not always sound has been 
occasionally recognised in the phenomenon of ``\idx{initial slip}'' 
\cite{Grad63,Haake83,Geigenmuller83,92d:82078}.  As developed in 
\cite{Roberts89b,Cox93b} a useful procedure for determining correct 
initial conditions for a low-dimensional model is based on the 
geometric picture of a centre manifold in the state space.

\subsection{Projecting initial conditions onto the model}
\label{SSprojic}

\begin{figure}[tbp]
	\centerline{\includegraphics{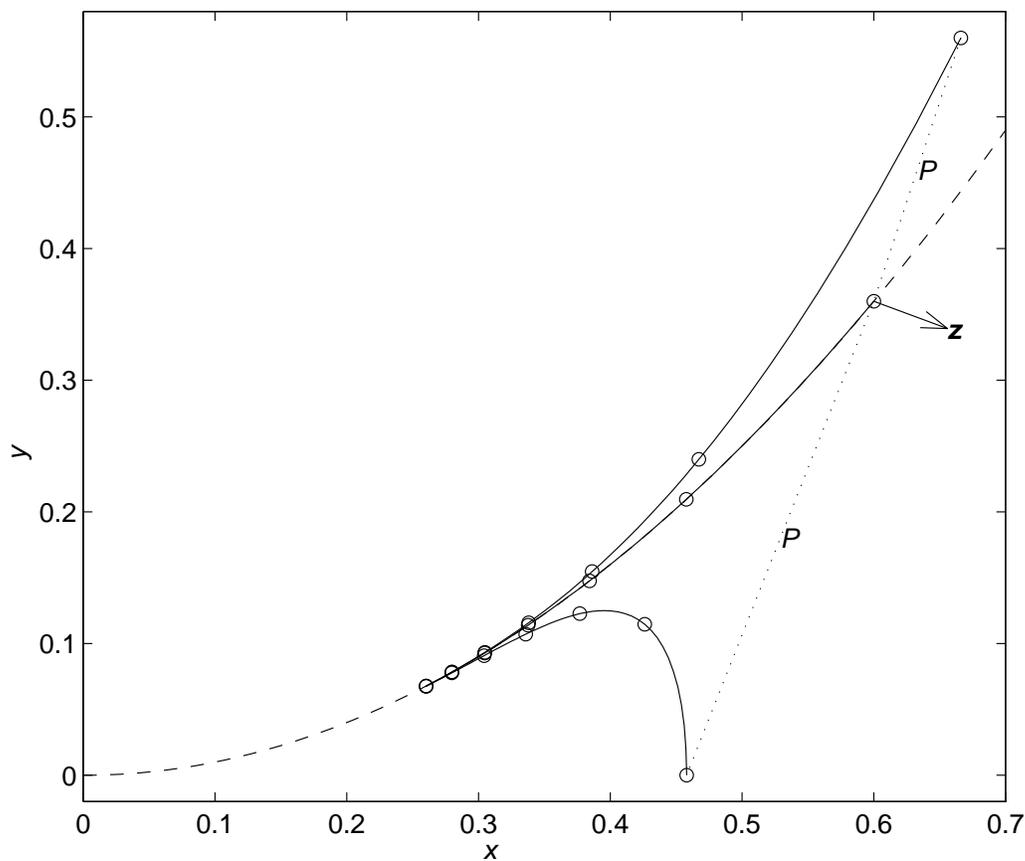}}
	\caption{trajectories (solid) of~(\protect\ref{Eegcm}) from three 
	different initial conditions, all with the same long-term dynamics 
	on the centre manifold $\cM_c$ (dashed) to an exponentially small 
	error.  The circles are plotted at $\Delta t=1$ apart.  The 
	modelling issue is to find the projection $P$ (dotted) from any 
	given initial condition off $\cM_c$ onto one for the model on 
	$\cM_c$.  $P$ is described by its normal vector $\vec z$ at 
	$\cM_c$.}
	\protect\label{Feginit}
\end{figure}%
The Relevance Theorem~\ref{Tcmrel}\index{relevance} 
assures us that there is indeed a particular solution of the 
low-dimensional model on $\cM_c$ which is approached exponentially by 
every trajectory of the \toe{} (provided the trajectory starts close 
enough to $\cM_c$).  This is illustrated in Figure~\ref{Feginit} where 
trajectories of the dynamical system~(\ref{Eegcm}) from three 
different initial conditions all have the same long-term evolution 
\emph{to a difference which decays exponentially quickly.} The 
modelling task is to find the \idx{projection} $P$ from any given initial 
state off $\cM_c$, say $\vec u_0$, onto a state on $\cM_c$, say $\vec 
v(\vec s_0)$, so that the long-term evolution, from $\vec s_0$ in the 
model, will be the same to an exponentially small error.  Some algebra 
finds this projection.

To be precise, the projection will be along the curved 
\emidx{isochronic manifolds}, or \emidx{isochrons} \cite{Winfree74,Guckenheimer75}, 
as shown in \cite{Roberts89b}.  However, as in the example plotted in 
Figure~\ref{Feginit}, over a large part of the state space a linear 
approximation to the projection may be quite adequate; ``linear'' in 
distance away from $\cM_c$, but varying with $\vec s$ along $\cM_c$.  
Such a linear projection is most easily defined by linearly 
independent normal vectors to the isochronic manifold at $\cM_c$, such 
as $\vec z= (1,-x/(1+2x^2))$ shown in Figure~\ref{Feginit}.  That is, 
we approximate by projecting along the tangent planes of the isochronic 
manifolds.  Then the \idx{initial condition}, $\vec s_0$, for the model is 
found from the requirement that the displacement is orthogonal to the 
normal vectors $\vec z_j$:
\begin{equation}
	\left\langle \vec z_j,\vec u_0-\vec v(\vec s_0)\right\rangle=0\,,
	\label{Eproj}
\end{equation}
where the angle brackets denote a suitable inner product.  For an 
$m$-di\-men\-sion\-al centre manifold within an $n$-di\-men\-sion\-al 
dynamical system, the isochronic manifolds are of dimension $n-m$ (the 
state space ``collapses'' by this many dimensions), and so we need $m$ 
linearly independent normal vectors $\vec z_j$ to define this projection.

In the immediate vicinity of the fixed point at the origin of the 
general dynamical system~(\ref{Egenfm}), linear arguments suggest the 
vectors $\vec z_j$ are eigenvectors, or generalised eigenvectors, of 
the adjoint eigen-problem
\begin{displaymath}
	\cL^\dag \vec z_j=\vec 0\,.
\end{displaymath}
These give the correct projection onto $\cE_c$ under the linear 
dynamics.  Under the nonlinear dynamics inherent upon the centre 
manifold $\cM_c$, dynamical arguments given in \cite{Roberts89b} show 
that to find the projection vectors $\vec z_j(\vec s)$ as a function 
of position on $\cM_c$, we solve
\begin{equation}
    \cD\vec z_j - 
    \sum_k\left\langle\cD\vec z_j,\vec e_k\right\rangle\vec z_k
    =\vec 0
    \quad\mbox{and}\quad
    \left\langle\vec z_j,\vec e_k\right\rangle=\delta_{jk}\,,
	\label{Eadj}
\end{equation}
where $\cD$ encapsulates the dynamics of trajectories near $\cM_c$ as 
\begin{displaymath}
		\cD{\vec z}=\D t{\vec z}+\cJ^\dag\vec z\,,
\end{displaymath}
in which the chain rule determines that
\begin{displaymath}
	\D t{\vec z}=\D{\vec s}{\vec z}[\cG\vec s+\vec g\left(\vec 
	s\right)]\,,
\end{displaymath}
and $\cJ^\dag$ is the adjoint of the Jacobian
\begin{displaymath}
	\cJ=\cL+\left.\D{\vec u}{\vec f}\right|_{\cM_c}\,.
\end{displaymath} 

As shown in \cite{Mercer90,Mercer94a,Watt94b}, upon applying the 
formulae~(\ref{Eproj}) and~(\ref{Eadj}) to the \idx{dispersion} of a 
contaminant in a channel or pipe, we accurately predict the different 
displacements of the mean concentration that occur from different 
release positions across the channel or pipe.  The similar phenomena 
of a long-term variance deficit in the spread of the contaminant is 
also predicted.  \emph{With the geometric picture of centre manifolds 
we have created a mechanism to provide correct initial conditions for 
low-dimensional dynamical models.}

\subsection{Normal forms also show the way}
\label{SSnorm}

Commonly, as intimated at the start of \S\ref{Sappl}, the normal 
form\index{normal forms} transformation is applied to the dynamics 
\emph{after} reduction from the \toe{} to the centre manifold.  The 
normal form then assisting in the classification of the dynamics.  
However, the normal form of the complete dynamics of the \toe{} 
clearly shows both the low-dimensional model (Elphick \etal{} 
\cite{Elphick87b}) and the correct projection of initial conditions 
(Cox \& I \cite{Cox93b}).

\begin{figure}[tbp]
	\centerline{\includegraphics{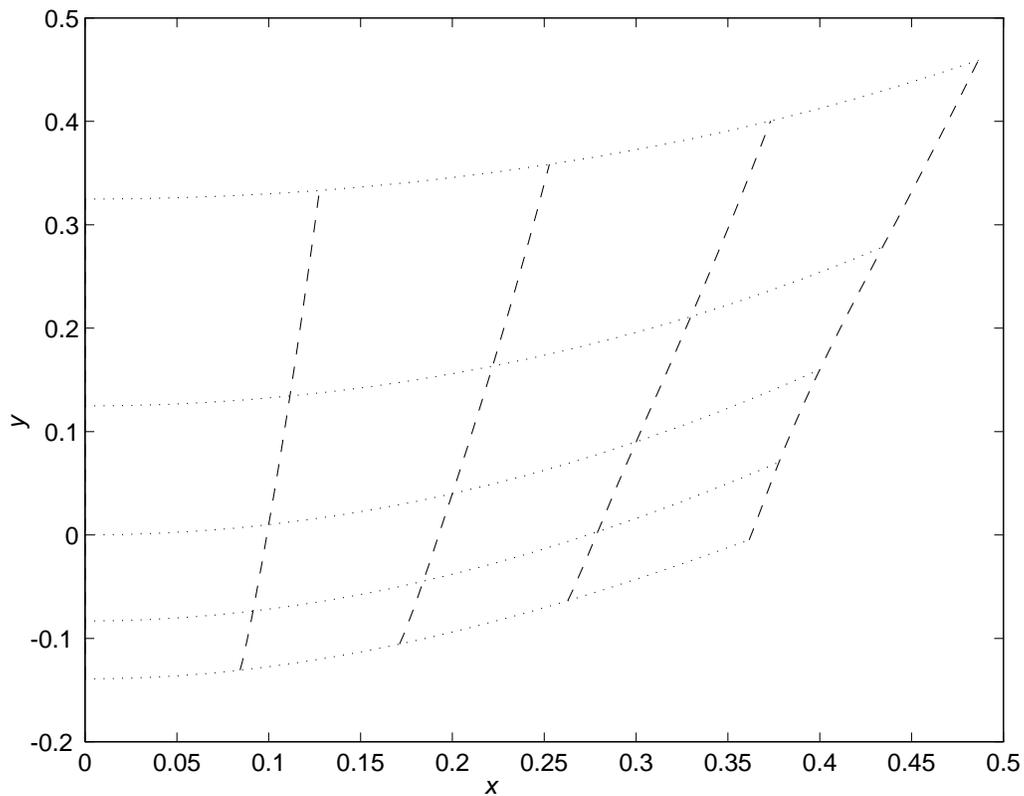}}
	\caption{the normal form transformation of~(\protect\ref{Eegcm}): 
	the dashed lines are the constant $X$ isochrons; the dotted lines 
	are constant $Y$; with $\Delta X=\Delta Y=0.1$.  The centre manifold 
	is the dotted curve ($Y=0$) emanating from the origin.}
	\protect\label{Fnorm}
\end{figure}

For simplicity of discussion suppose that the dynamics of the \toe{} 
are given in the \idx{separated form}~(\ref{Estdfmx}--\ref{Estdfmy}).  Then 
we may seek a nonlinear, but near identity, coordinate 
transformation
\begin{displaymath}
	\vec x=\vec X+\vec \Xi(\vec X,\vec Y)\,,
	\quad\mbox{and}\quad
	\vec y=\vec Y+\vec\Psi(\vec X,\vec Y)\,,
\end{displaymath}
such as that shown in Figure~\ref{Fnorm}, so that the \toe{} is 
transformed to
\begin{equation}
	\begin{array}{rcl}
		\dot{\vec x} & = & A\vec x+\vec f(\vec x,\vec y)  \\
		\dot{\vec y} & = & B\vec y+\vec g(\vec x,\vec y)
	\end{array}
	\longrightarrow
	\begin{array}{rcl}
		\dot{\vec X} & = & A\vec X+\vec F(\vec X)  \\
		\dot{\vec Y} & = & B\vec Y+G(\vec X,\vec Y)\vec Y
	\end{array}
	\label{Enormf}
\end{equation}
This is always possible given the pattern of eigenvalues for the 
existence of a centre manifold.  For the example system~(\ref{Eegcm}), 
the nonlinear coordinate transformation
\begin{eqnarray*}
	x & = & X+XY+\frac{3}{2}XY^2+\cdots  \\
	y & = & Y+X^2+2Y^2+4Y^3+\cdots\,,
\end{eqnarray*}
shown in Figure~\ref{Fnorm}, separates the dynamics to
\begin{displaymath}
	\dot X=-X^3\,,
	\quad\mbox{and}\quad
	\dot Y=-Y-2X^2Y+\cdots\,.
\end{displaymath}
In the form~(\ref{Enormf}), observe that $\vec Y=\vec 0$ is clearly 
the invariant and exponentially attractive centre manifold.  But also, 
both on and off $\cM_c$, the $\vec X$ evolution is completely 
independent of the decaying modes $\vec Y$.  Hence all solutions of 
the \toe{} from initial conditions with the same $\vec X$ have 
precisely the same $\vec X$ evolution for all time, and thus they all 
tend to the same solution on the centre manifold.  Thus the normal 
form shows that the projection of initial conditions\index{initial 
condition} onto the centre manifold is along constant $\vec X$.  
Constant $\vec X$ are the \idx{isochronic manifolds}.

The normal form transformation works its magic by decoupling the 
centre modes, $\vec X$, from the quickly decaying modes $\vec Y$.  For 
\idx{slow manifold} models which neglect fast oscillations, that is 
\idx{guiding centre} models, a normal form coordinate transform will 
similarly exhibit the slow manifold as $\vec Y=\vec 0$ 
\cite{Cox93a,Cox93b}.  The coordinate transform will also remove terms 
linear in $\vec Y$ from $\dot{\vec X}$.  However, in general, terms 
quadratic in $\vec Y$ cannot be eliminated from $\dot{\vec X}$---there 
is generally a resonant forcing\index{resonance} of the slow modes, 
$\vec X$, by the fast oscillations or waves, $\vec Y$, which is 
quadratic in the oscillation amplitude.  Thus, as mentioned in 
\S\ref{Sslow}, the evolution of the \toe{} is generally slightly but 
unavoidably different to that of the slow model.  An example would be 
the phenomenon of Stokes drift generated by any ``fast'' water waves 
superimposed upon the dynamics of large-scale currents.  There are 
limitations to low-dimensional models based upon the concept of a slow 
manifold (or a \idx{sub-centre manifold}).

Although the normal form of the \toe{} is a very useful conceptual 
tool, it does not provide a practical method for constructing a 
low-dimensional model.  The reason is that it involves considerable 
wasted algebra in the transformation of the ultimately neglected 
stable and/or fast modes.  If there are many stable mode, as usual in 
interesting physical systems, then the normal form transformation may 
be practically impossible, whereas methods based upon the 
Approximation Theorem~\ref{Tcmapprox} will be manageable.

\section{Enforcing some surprises}
\label{Sforc}

So far we have implicitly restricted attention to unforced dynamical 
systems.  The presence of small \idx{forcing} in the \toe{} may be 
transformed into a forcing of the model.  Discussed in the following 
subsections are the cases of deterministic forcing and of stochastic 
forcing or noise.  The centre manifold formalism, coupled with the 
projection of initial conditions, permits an accurate modelling of the 
effects of the forcing.  Remarkably, small forcing that otherwise 
would be neglected can have a large effect on the model's dynamics.

\subsection{Deterministic forcing}
\label{SSdforc}

Consider the dynamical system~(\ref{Eegcm}) with the $y$ mode 
forced by a steady effect $\epsilon$:\index{forcing, constant}
\begin{equation}
	\dot x= -xy 
	\quad\mbox{and}\quad
	\dot y=-y+x^2-2y^2 \,-\epsilon\,.
	\label{Eegcmfor}
\end{equation}
As for the initial condition problem, a simple projection of the 
forcing onto $\cE_c$ would indicate that this particular forcing would 
have little influence on the low-dimensional model.  However, this is not 
so.  The correct model is 
that on the perturbed centre manifold $y\approx 
-\epsilon+(1+2\epsilon)x^2$ the evolution is
\begin{equation}
	\dot x\approx \epsilon x-(1+2\epsilon)x^3\,,
	\label{Eegfcm}
\end{equation}
which exhibits a \idx{pitchfork bifurcation} with two stable fixed 
points at $x\approx\sqrt\epsilon$.  This is a \idx{large response} in 
the model to a small forcing.  Such a significant change in a model 
may be typical because the slow evolution on the centre manifold is 
sensitive to small influences.

Forcing may be incorporated into a model by two approaches.  Firstly, 
if \emph{the forcing is constant}, the \toe{} is autonomous, then a 
simple trick will suffice to put the dynamical system within the 
rigorous scope of centre manifold theory.  For the above example, 
simply substitute $\epsilon=\delta^2$, say, and adjoin the trivial 
dynamical equation $\dot\delta=0$.  Then the forcing becomes a 
nonlinear term in the dynamics, the spectrum shows a centre manifold 
parameterised by the modes $x$ and $\delta$, and the construction of 
the centre manifold and the evolution thereon leads to the forced 
model~(\ref{Eegfcm}).
\begin{figure}[tbp]
	\centerline{\includegraphics{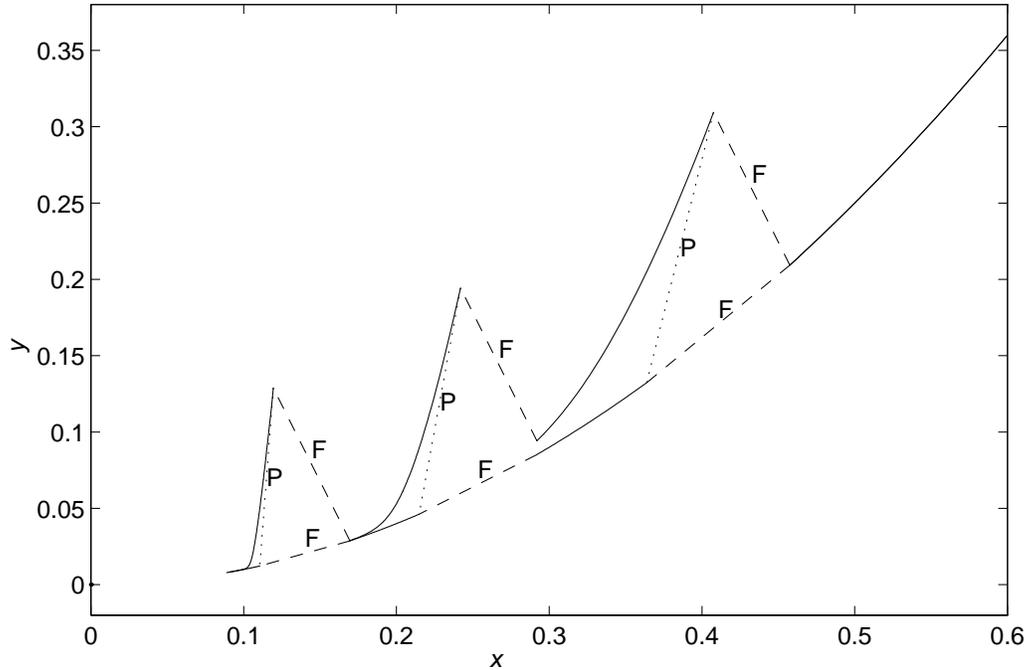}}
	\caption{evolution of a dynamical system with discrete impulsive 
	forcing and the corresponding evolution of its low-dimensional 
	model on $\cM_c$.  The trajectories (solid) are ``kicked'' by each 
	impulse (\textsf{F}, dashed).  To maintain the correspondence 
	between the exact solution and its model, the impulse has to be 
	projected (\textsf{P}, dotted) along the isochronic manifolds onto 
	$\cM_c$ to become an impulsive forcing for the model.}
	\protect\label{Fforc}
\end{figure}

Secondly, \emph{for time-dependent forcing}\index{forcing, 
time-dependent} an argument developed in \cite{Roberts89b} will 
suffice.  Suppose a general dynamical system~(\ref{Egenfm}) has a 
small applied forcing $\vec F(t)$, namely
\begin{equation}
	\dot{\vec u}=\cL\vec u+\vec f(\vec u)+\vec F(t)\,.
	\label{Egenfor}
\end{equation}
Imagine approximating the forcing by a sum of \idx{impulses}, $\vec 
F=\sum_j\vec F_j\delta(t-t_j)$, at discrete times $t_j$.  Then within 
each force free interval, the centre manifold is exponentially 
attractive, as shown in Figure~\ref{Fforc}, and the unforced model 
applies.  Each impulse ``kicks'' the dynamical system off $\cM_c$, see 
Figure~\ref{Fforc}.  As in the \idx{initial condition} problem, the 
evolution following such a ``kick'' exponentially quickly approaches 
the evolution of some particular solution on $\cM_c$.  This solution 
is also obtained by some particular impulse, found by projecting the 
impulse $\vec F_j$, applied to the low-dimensional model.  The 
sequence of such projected impulses then approximates a continuous 
forcing of the model, say $\vec G(\vec s,t)$ in
\begin{equation}
	\dot{\vec s}=\cG\vec s+\vec g\left(\vec s\right)+\vec G(\vec s,t)\,.
\end{equation}
Letting
\begin{displaymath}
	B_{ij}=\mbox{inverse of }\left\langle\vec z_i,\D{s_j}{\vec 
	v}\right\rangle\,,
\end{displaymath}
the appropriate projection of the forcing of the \toe{} into a forcing 
of the model is
\begin{equation}
	G_i=\sum_j B_{ij}\left\langle\vec z_j,{\vec	F}\right\rangle\,.
\end{equation}
This argument is accurate to errors $\Ord{F^2}$, it gives the linear 
projection of the forcing.  Algebraic formula for higher order 
corrections, in $F$, have not yet been found.

For the example~(\ref{Eegcmfor}), $G=\epsilon x$ and hence the forced 
model is
\begin{equation}
	\dot x\approx \epsilon x-x^3\,.
\end{equation}
This model is apparently different to the earlier~(\ref{Eegfcm}).  The 
reason for the difference is that \emph{implicitly} we have used a 
different parameterisation of the forced centre manifold.  Cox \& I 
\cite{Cox91} showed that there is extra freedom in parameterising a 
forcing because such forcing typically moves the evolution off the 
centre manifold into new regions of state space.  However, we also 
showed that, in general, the forcing in the model then undesirably 
depends upon an integral over the previous history of the time-dependent 
forcing---a \idx{memory} effect as also noted in \S\S\ref{SSdisp} for shear 
dispersion \cite{Smith83b}.  The \emph{only} way to eliminate memory 
integrals in the dynamics of the model is to use the \idx{isochronic 
manifolds} as a basis for the parameterisation off $\cM_c$, as done 
implicitly in the argument above.  \emph{The geometric picture 
provided by centre manifold theory enables us to correctly incorporate 
forcing.}

In application to beam theory\index{elastic beam} \cite{Roberts93}, 
for example, the correct treatment of forcing allows us to predict the 
``toothpaste effect'', that is, we predict the longitudinal extrusion 
of an elastic rod which results from a purely lateral compression!

\subsection{Stochastic dynamical systems}
\label{SSnoise}

In practice we may need to model dynamics in a noisy\index{noise} 
environment \cite[e.g.]{91a:82034}.  Just one example of interest is 
the \idx{dispersion} of a contaminant in a turbulent river---the 
\idx{turbulence} can only reasonably be modelled by invoking 
stochastic factors.  How do we construct low-dimensional models of 
\idx{stochastic differential equations} (\sde{}s) using centre 
manifold ideas and techniques?

Currently there seem to be a number of largely disparate threads in 
the modelling of \sde{}s.  Boxler, Arnold \etal{} 
\cite{Boxler89,92e:58184,94i:60066} prove theoretical results about 
low-dimensional centre manifolds of \sde{}s.  However, the analysis is 
sophisticated and the results, as seen in examples, seem difficult to 
apply.  Analysis of \idx{Fokker-Plank equation}s lead to ``\idx{weak 
models}'' \cite{Knobloch83,88i:76018,89c:58087,95b:58106}, ``weak'' in 
the sense that detailed information about the noise is irretrievably 
lost to the model.  Systematic \idx{adiabatic elimination} 
\cite{Schoner86,Schoner87,88h:58087} and normal form\index{normal 
forms} transformations 
\cite{Coullet85,Srinamachchivaya90,Srinamachchivaya91} have also been 
suggested, but typically suffer from introducing many complicated 
noise processes into the dynamical model.

Inspired by the work on forcing described in the previous section, 
Chao \& I \cite[\S3]{Chao95} realised that much of the complication 
may be removed from the noise in a low-dimensional model.  On the long 
time-scale of the evolution on the centre manifold, the noise 
processes are essentially white and a little algebra, based upon the 
freedom to parameterise the forced centre manifold, explicitly shows 
this.  However, such simplification only works for effects linear in 
the stochastic noise.  Effects quadratic in the noise are more 
complicated.  For example, stochastic resonance\index{resonance, 
stochastic} leads to \emidx{irreducible noise} processes appearing in 
the model, such as $w(t)$ from the (Stratonovich) \sde{}
\begin{equation}
	dw = Z\,dW\,,  
	\quad\mbox{and}\quad
	dZ = -\beta Z\,dt+dW\,,
	\label{Eresnoi}
\end{equation}
for a given noise $W(t)$ in the \toe.  Observe that $Z$ is a coloured 
noise generated directly by $W$, then $Z$ and $W$ together induce the 
new noise $w$.  Neither normal form nor centre manifold algebra can 
simplify the representation of the induced stochastic process $w(t)$.  
However, a centre manifold analysis of the corresponding \idx{Fokker-Plank 
equation} \cite[\S4]{Chao95} suggests that we replace $w(t)$ by its 
long-term drift and fluctuation,
\begin{equation}
	dw\approx\frac{1}{2}dt+\frac{1}{2\sqrt{\beta}}dW'\,,
	\label{Eeffnoi}
\end{equation}
where $W'(t)$ denotes a \emph{new} and \emph{independent} noise.  In 
essence, the nonlinear machinations in~(\ref{Eresnoi}) extracts 
information about $W$ that cannot be obtained by sampling $W$ over 
large times, information which is thus independent and can only be 
represented on the long time-scales of the model by a new noise 
process.

Much further research is needed to apply these ideas and provide 
theoretical support.

\section{At boundaries}
\label{Sbound}

Models expressed as partial differential equations, such as those for 
\idx{dispersion} in a river and for beam theory,\index{elastic beam} 
require \idx{boundary conditions} at the limits of the domain, the 
inlet and outlet \cite{Smith88} or the ends of the beam respectively.  
Such models as these are typically derived through a slowly-varying 
approximation under the assumption that the domain of interest is 
arbitrarily large as discussed in \S\ref{Sappl}.  However, typical 
physical situations of interest possess \idx{finite domains}.  The 
issue is: what are the correct boundary conditions to be used at the 
edge of the domain for such model equations?  The provision of correct 
boundary conditions are sometimes crucial---in large aspect-ratio 
\idx{convection} the boundary drives the entire nature of the 
long-term dynamics, see \cite[p444]{Newell93} or 
\cite[\S3]{Greenside96}.

Relatively little work has been done on determining correct boundary 
conditions.  When interested in generic dynamical behaviour one 
typically just uses periodic boundary conditions, 
\cite{Kuramoto78,Hyman86a,Armbruster89} for example.  Alternatively, 
one may only allow boundaries in the original problem which will fit 
neatly into the scheme of the asymptotic approximation, 
\cite{Chapman80a,Segel69,Roberts93} for example.  These two choices 
are forced by the inability of primitive asymptotic schemes, such as 
the method of \idx{multiple scales}, to embrace the presence of typical 
physical boundaries.  Alternatively, in realistic physical problems 
heuristic arguments may supply approximate boundary conditions for the 
model, but are they mathematically justifiable?  Examples are the 
boundary conditions used with beam theory for the idealisations of 
free, fixed or pinned ends.  Cross and others 
\cite{Cross82,Cross83,Martel96} have used matched 
asymptotics to derive boundary conditions for two-dimensional 
convective rolls; however, their analysis is linear near the boundary, 
and yet needs to be nonlinear in the interior.  More attention needs 
to be given to this important aspect of modelling.

Some special invariant manifolds (\S\S\ref{SSrat}) in conjunction with 
the correct projection of initial conditions (\S\ref{Sinitial}) 
provide a route to determine correct boundary conditions 
\cite{Roberts92c}.  Boundary conditions are provided by two separate 
and complementary arguments; both relying on the same trick of 
investigating the \emidx{spatial evolution} away from the boundary 
into the interior (also mentioned in \S\ref{Sappl}), given also a slow 
time evolution.  In essence, we probe the dynamics of the 
\idx{boundary layers} at either end of the domain and how they merge 
into the slowly-varying interior solution.  The dominant terms in the 
boundary conditions typically agree with those obtained through simple 
physical arguments.  However, refined models of higher order require 
subtle corrections to the previously-deduced boundary conditions, and 
also require the provision of additional boundary conditions to 
eliminate unphysical predictions and to form a complete model.

\subsection{The physical boundary layer}

In the {\toe} with slow time evolution, the \idx{spatial evolution} 
away from a boundary consists of: exponentially decaying ``stable'' 
modes; exponentially growing modes which are not seen as the far 
distant boundary removes them; and nearly neutral modes which 
correspond to those of the interior model.  Thus near any boundary, 
because of the dynamics inherent in the differential equations of the 
\toe{}, the system must lie in the \idx{centre-stable manifold} (to an 
exponentially small error in the length $L$ of the domain).  This 
centre-stable manifold may be constructed.

As an illustration of the ideas, consider the shear \idx{dispersion} problem 
shown in Figure~\ref{Fchan}.  If the time dependence is negligible, 
solutions $c'(y)e^{\lambda x}$ exist where
\begin{equation}
	\frac{d^2c'}{dy^2}+(\delta\lambda^2-u(y)\lambda)c'\approx 0\,.
	\label{Espat}
\end{equation}
There exists solutions:
\begin{itemize}
	\item  $\lambda=0$, $c'$ constant, corresponding to the critical mode 
	of the slowly-varying centre manifold model;

	\item  modes with negative eigenvalues, approximately $\lambda=-3.414$, 
	$\lambda=-12.25$, etc, describing how the cross-stream diffusion, 
	$d^2c'/dy^2$, exponentially smooths out details of the inlet 
	concentration as it is advected downstream by the profile $u(y)$;

	\item and modes with large positive eigenvalues, $\lambda\propto 
	1/\delta$, allowing upstream diffusion to accommodate the interior 
	concentration to any particular outlet condition.
\end{itemize}
The centre-stable manifold near the inlet is approximately the space 
spanned by the critical mode with $\lambda=0$ together with the modes 
of negative $\lambda$.  Conversely, at the outlet and viewing the 
dynamics towards negative $x$, the centre-stable manifold is 
approximately the span of the critical mode together with the modes of 
positive $\lambda$.

A boundary condition must be given as part of the \toe{}.  Such a 
boundary condition intersects the constructed centre-stable manifold 
to form the set of possible states that may hold at the boundary.  
These states form a set of possible ``initial conditions'' for the 
spatial evolution away from the boundary.  The states in the 
intersection are then projected onto the centre manifold (of the 
spatial evolution) using the \idx{initial condition} arguments 
described in \S\ref{Sinitial} to give a set of allowable states 
for the centre manifold model at the boundary.  These determine some 
boundary conditions for the low-dimensional model.

For the example of shear dispersion in a pipe or channel, these 
arguments generally provide one inlet condition for the \idx{Taylor 
model}~(\ref{Echantay}), but provide no constraint at the outlet 
\cite[\S2,\S3.1]{Roberts92c} because the centre-stable manifold at the 
outlet is of one higher dimension, thus has a larger intersection 
with the exit boundary conditions, and hence does not restrict the 
states of the model at the outlet.

\subsection{The boundary layer of the model}

As noted for Taylor's model of dispersion, some low-dimensional models 
need more boundary conditions than the previous argument supplies.  
Such boundary conditions come from considering the dynamics of the 
model in the \idx{boundary layers} at either end of the domain.

Given slow variations in time, the \idx{spatial evolution} of the 
\emph{model} into the interior consists of: critical modes which 
correspond to those of the low-dimensional model seen in the interior; 
unphysical exponentially decaying modes which do not penetrate far 
into the interior; and unphysical exponentially growing modes which 
the far boundary must remove.  We require that there be no unphysical 
stable modes near a boundary \cite[\S3.2]{Roberts92c} as otherwise 
there will be unsightly and unphysical transients in the model's 
solutions.  Thus the system must lie in the \idx{centre-unstable 
manifold} of the model, which we may construct.  The fact that any 
solution actually lies in the centre manifold is assured by the far 
boundary because a boundary's unstable modes correspond to the far 
boundary's removed stable modes.  This requirement provides sufficient 
additional boundary conditions \cite[\S3.3]{Roberts92c}.

For example, in the \idx{Taylor 
model}\index{dispersion}~(\ref{Echantay}) and for negligible time 
variations, solutions $e^{\lambda x}$ exist with $\lambda\approx 0$ 
and $\lambda\approx U/D$.  At the outlet, the stable mode, 
approximately $e^{Ux/D}$, is removed by enforcing a condition that the 
system is on the centre-unstable manifold (here just the centre 
manifold as there is no unstable mode).  This provides a boundary 
condition at the exit to complement the entry boundary condition from 
the previous argument.  The argument here does not provide an entry 
boundary condition as there is no stable mode in the Taylor model at 
the inlet.

These two arguments together very neatly provide boundary conditions 
for models with one slowly-varying spatial dimension and slow 
variations in time.  Outstanding is a resolution of the apparent 
differences between this approach and the more specialised approach 
employed for beams in \cite{Roberts93}, and the provision of boundary 
conditions for oscillatory critical modes.  Also outstanding, but of 
considerable interest, is the issue of what to do for models involving 
\emph{two} spatial dimensions, such as models of thin, elastic plates 
and of planform evolution in convection.

\section{Computer algebra handles the details}
\label{Scompalg}

Computing details of the shape of a centre manifold and the evolution 
thereon may involve copious algebra.  A leading order approximation is 
often tractable by hand, but when this approximation is not 
structurally stable then the requisite higher-order\index{higher order 
models} corrections are typically tedious to determine.  Extremely 
high-order calculations, for example to determine the spatial 
resolution of dispersion \S\S\ref{SSdisp} or beam theory 
\cite{Roberts93}, are grossly tedious.  Computer 
algebra\index{computer algebra} may reduce the human labour involved.  
The challenge is to develop algorithms which are simple to reliably 
implement.

\subsection{Explicit series expansions}

Various \idx{computer algebra packages} have been used to compute centre 
manifolds, such as \textsc{Maple} \cite{95j:34100}, \textsc{Reduce} 
\cite{90k:58151} and \textsc{macsyma} \cite{Rand85}, and also normal forms 
\cite{Rand85,91i:65148}.

In problems with centre manifolds of small dimension, 1, 2 or 3 say, 
the usual approach is simply to seek an explicit multinomial 
form.\index{multivariate expansion} One then substitutes into the 
governing equations, gathers like terms, and solves for the unknown 
coefficients.  See for example Rand \& Armbruster's 
\cite[pp27--34]{Rand87} \textsc{Macsyma} code or Freire \etal{}'s 
\cite{90k:58151} \textsc{Reduce} code for constructing the centre 
manifold of a finite-dimensional dynamical system.  But this approach 
fails for centre manifolds of higher dimension, especially for the 
``infinite'' dimensional centre manifolds associated with 
spatio-temporal models.  For example, in lubrication models of thin 
film\index{thin films} flow (\S\ref{SSfilm}) there are 5 basic terms 
of the leading, fourth-order, namely $\eta_x^4$, $\eta_x^2\eta_{xx}$, 
$\eta_{xx}^2$, $\eta_x\eta_{xxx}$ and $\eta_{xxxx}$, each of these 
possibly combined with the multiplication by an arbitrary power of 
$\eta$.  \emph{It very soon becomes impractical to code a complete sum 
of general terms.}

Another approach is to select one or two ``ordering parameters'' that 
measure the size of all nonlinearities and all other small 
effects.\index{small parameters} Then express the model explicitly in 
terms of asymptotic sums in these ordering parameters with functional 
coefficients which are to be determined \cite[e.g.]{Roberts94c,Mei95}; 
the form of the unknown coefficients are found in the working, a 
general form need not be explicitly written down beforehand.  To 
reduce the dynamics onto the centre manifold, one then has to 
substitute the asymptotic sums into the governing equations, reorder 
the summations, rearrange to extract dominant terms, and evaluate the 
expressions.  While perfectly acceptable when done correctly, it leads 
to formidable working which obscures the construction of a model.  
Further, such asymptotic expansions, in common with the method of 
\idx{multiple scales}, reinforces the notion that careful balancing of 
the ``order'' of small effects are necessary in the 
\emph{construction} of a model rather than in its \emph{use} in some 
situation (\S\ref{Ssmall}).

\subsection{Implicit approximation}

\begin{figure}[tbp]
\centerline{{\tt    \setlength{\unitlength}{1.1pt}
\begin{picture}(288,179)
\thinlines    \put(25,163){\line(5,-3){155}}
              \put(73,160){\res}
              \put(250,54){$\vec\phi,\vec\psi$}
\multiput(179,70)(0,-5){9}{\circle*{1}}
\put(179,24){\line(-5,3){76}}
\multiput(103,70)(0,5){8}{\circle*{1}}
\put(103,107){\line(5,-3){62}}
\put(135,70){\circle*{4}}
\put(136,71){$\cM_c$}
\put(136,98){$\cL$}
\thicklines   
\put(10,160){\line(5,-1){20}}
\put(30,156){\line(5,-2){20}}
\put(50,148){\line(5,-3){20}}
\put(70,136){\line(5,-4){20}}
\put(90,120){\line(1,-1){20}}
\put(110,100){\line(5,-6){40}}
\put(150,52){\line(1,-1){20}}
\put(170,32){\line(5,-4){20}}
              \put(65,10){\vector(0,1){159}}
              \put(5,70){\vector(1,0){263}}
\end{picture}}
} 
	\caption{schematic diagram of an iteration to determine the centre 
	manifold.  It depicts the linear operator, $\cL$, at the fixed point 
	at the origin guiding corrections to the description, $\vec v$ \& 
	$\vec G$, based upon the residuals of the \toe.}
	\protect\label{Fcmit}
\end{figure}
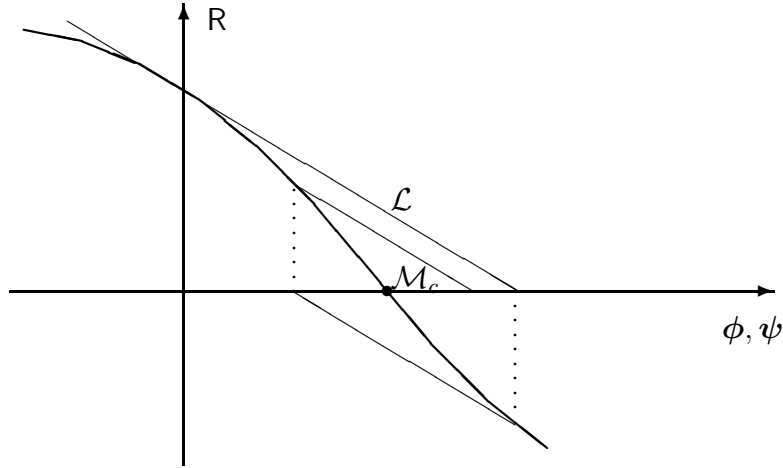

Recently I proposed \cite{Roberts96a} an iterative method, based upon 
the \idx{residual}s of the governing differential equations of the 
\toe, as shown schematically in Figure~\ref{Fcmit}.  The evaluation of 
the residuals is a routine algebraic task which may be easily done 
using computer algebra by simply coding the governing differential 
equations; it replaces the whole messy detail of the manipulation of 
asymptotic expansions (e.g.~\cite[\S5.4]{Coullet83}).  In the same 
spirit as comments made by Barton \& Fitch \cite[\S2.3]{Barton72} in 
1972, the aim of this proposed approach is to \idx{minimise human 
time} by using a novel algorithm which is simply and reliably 
implemented in computer algebra, albeit with inefficiencies in the use 
of computer resources.

Given a generic dynamical system~(\ref{Egenfm}) we seek a centre 
manifold $\vec u=\vec v(\vec s)$ on which $\dot{\vec s}=\cG\vec s+\vec 
g(\vec s)$.  As in \S\S\S\ref{SSSapprox}, if $\vec\phi$ and $\vec\psi$ 
approximate $\vec v$ and $\dot{\vec s}$ respectively, then I argue 
\cite{Roberts96a} that corrections, $\vec\phi'$ and $\vec\psi'$, may 
be found by solving the \emidx{homological equation}
\begin{equation}
	\cL\vec\phi'-\D{\vec s}{\vec\phi'}\cG\vec s -\cV\vec\psi'
	=\res(\vec\phi,\vec\psi)\,,
	\label{Eresit}
\end{equation}
where $\cV=\partial\vec\phi/\partial\vec s|_{\vec {0}}$, and recall 
that $\res$ is the residual of the \toe.  Generally this produces 
linear convergence to $\cM_c$ and the low-dimensional evolution, 
linear in the sense that each iteration corrects the next order or two 
of the expressions.  The algorithm for the computer algebra derivation 
of low-dimensional models is relatively: simple to implement, because 
the computation of the residual is via a \emph{direct} coding of the 
governing differential equations; flexible because many changes to the 
\toe{} can be simply accommodated by updating the residual 
computation; and \idx{reliable}, because it relies upon the actual residual 
going to zero---any error is picked up by a lack of convergence.

This algorithm has already been used to develop models of thin 
film\index{thin films} fluid dynamics upon curved substrates with 
gravitational and inertial effects \cite{Roberts96b,Roy96}.  A 
currently outstanding issue is developing a companion iteration scheme 
to determine the correct projection of \idx{initial condition}s 
(\S\ref{Sinitial}) and \idx{forcing} (\S\ref{Sforc}).

\section{Conclusion}

We have looked at geometric based arguments about how to form accurate and 
reliable low-dimensional models of dynamical systems.  The aim is to 
take a so-called ``Theory Of Everything'' ({\toe}), a complete but far 
too detailed description of the system, and produce a ``coarse 
grained'' model that captures all the dynamics, and little else, that 
are of interest in a particular application.  Although the geometric 
arguments are supported centre manifold theory, we have seen that many 
applications go beyond the current rigorous support.  

This approach to modelling has many advantages:
\begin{itemize}
\item the signature of the model is straightforwardly derived from 
linearisation (\S\S\S\ref{SSSexist}, \S\ref{Sslow}), but one can be 
inventive (\S\S\ref{SSextend}, \S\S\ref{SSfilm}, \S\S\ref{SSband});

\item the model may be systematically refined (\S\S\S\ref{SSSapprox}, 
\S\S\ref{SSextend}, \S\S\ref{SSdisp}), perhaps using computer algebra 
to handle most of the details (\S\ref{Scompalg});

\item one may be very flexible about the relative magnitude of 
``small'' parameters, resulting in a model which is later 
justifiably tuned to specific applications (\S\ref{Ssmall});

\item it provides correct initial conditions for forecasts 
(\S\ref{Sinitial}), initial conditions that guarantee fidelity between 
the predictions of the model and the long-term evolution of the full 
dynamics;

\item may account for smooth or stochastic forcing (\S\ref{Sforc}); 

\item establishes an approach to determine correct boundary conditions 
for the models (\S\ref{Sbound}).
\end{itemize}

How to decide upon an accurate {\toe} on which to base the analysis is 
another problem---a model is only guaranteed if it is based upon an 
accurate description of the dynamics.

\addcontentsline{toc}{section}{References}
\markright{\textnormal{\sf References}}
{\raggedright

}

\newpage
\addcontentsline{toc}{section}{Index}
\markright{\textnormal{\sf Index}}
\input{lowd.ind}

\end{document}